\pdfoutput=1
\documentclass[letterpaper,journal]{IEEEtran}
\usepackage{orcidlink}

\usepackage{graphicx} 
\usepackage{algpseudocode}
\usepackage{amssymb}
\usepackage{amsmath,amsfonts}
\usepackage{xcolor}
\usepackage{graphicx}
\usepackage[utf8]{inputenc}
\usepackage[T1]{fontenc}
\usepackage{hyperref}
\usepackage{blindtext}
\usepackage{array}
\usepackage{enumitem}
\usepackage{pifont}
\usepackage{mismath}
\usepackage{multirow}
\usepackage{booktabs}
\usepackage[export]{adjustbox} 
\usepackage{mdframed}
\usepackage{makecell}

\usepackage{subcaption}

\usepackage[abbreviations]{foreign}

\usepackage{pifont} 
\usepackage{xcolor}
\usepackage{comment}

\usepackage{xspace}
\newcommand{\bingo}{\textnormal{\textsc{Bingo}}\xspace}
\newcommand{\avss}{\textnormal{\textsc{SimplAVSS}}\xspace}
\newcommand{\light}{\textnormal{\textsc{LightAVSS}}\xspace}
\newcommand{\hbacss}{\textnormal{\textsc{hbACSS}}\xspace}
\newcommand{\honey}{\textnormal{\textsc{HoneyB}}\xspace}
\newcommand{\haven}{\textnormal{\textsc{Haven}}\xspace}
\newcommand{\goss}{\textnormal{\textsc{GoSSamer}}\xspace}

\newcommand{\afourthofaline}{0.245\linewidth}
\newcommand{\halfaline}{0.49\linewidth}

\newcolumntype{P}[1]{>{\centering\arraybackslash}m{#1}}

\begin{document}

\title{Asynchronous Verifiable and Complete Secret Sharing: From Design Patterns to System-Level Trade-offs}

\author{Joachim Bruneau-Queyreix$^{1}$,
    Martin Zivojinovic$^{2}$,
    Leo Mendiboure$^{3}$,
    Tom Claudel$^{1}$,
    Laurent Réveillère$^{1}$
    \thanks{$^{1}$Univ. Bordeaux, CNRS, Bordeaux INP, LaBRI, UMR 5800, F-33400 Talence, France}
    \thanks{$^{2}$KTH Royal Institute of Technology, Stockholm, Sweden}
    \thanks{$^{3}$LIUPPA, Université de Pau et des Pays d'Adour, F-64600, Anglet, France.}
}

\maketitle

\begin{abstract}
    Modern secret sharing protocols promise robust secret sharing in fully asynchronous Byzantine systems, but their practical costs remain difficult to compare across papers, assumptions, and workloads.
    The difficulty is not merely experimental: protocols with similar asymptotic guarantees often place very different work on the critical path through their choices of dissemination, verification, batching, and recovery.
    This paper turns that fragmented research landscape into a corpus-backed design-space analysis and systems evaluation of asynchronous verifiable and complete secret sharing.
    We derive a design sieve, select seven representative protocols, and implement them in a common Rust framework.
    In our experiments on the Grid'5000 platform, we evaluate honest-dealer and faulty-dealer stress executions using latency, throughput, message count, bandwidth, and aggregate CPU-time.
    The results show that message complexity alone poorly predicts latency, that optimistic paths can be dominated by recovery cost, and that batching or packing helps only when verification stays off the critical path.
    These findings turn protocol-by-protocol comparisons into design-to-cost guidance for choosing, implementing, and extending asynchronous verifiable and complete sharing protocols.
\end{abstract}

\section{Introduction}

Secret sharing is a fundamental primitive in distributed cryptography, enabling a dealer to split a secret into shares distributed among $n$ parties, so that only authorized groups can reconstruct it.
In a $(t,n)$-threshold secret sharing scheme such as Shamir's~\cite{shamir1979share}, a dealer encodes a secret $s$ as the constant term of a random polynomial of degree $t-1$ over a finite field and distributes a distinct polynomial evaluation as a share to each of the $n$ parties.
Any $t$ shares reconstruct $s$ via Lagrange interpolation, while any $t-1$ shares reveal nothing.
Verifiable Secret Sharing (VSS~\cite{feldman1987practical, chor1985verifiable, rabin1989verifiable}) augments this by adding verification data, typically commitments and authentication, so that each party can check that its share is consistent with a single underlying secret and that the dealer did not misbehave.
If not, the misbehavior is detected and the protocol aborts or triggers share recovery.
For instance, Feldman's VSS~\cite{feldman1987practical} commits to the polynomial coefficients, enabling local share verification against these commitments.

VSS is a core building block for higher-level protocols, including secure multiparty computation, distributed threshold encryption~\cite{christodorescu2021amortized}, secure and private sorting~\cite{agarwal2024secure}, Byzantine agreement~\cite{liu2018scalable, vassantlal2022cobra}, distributed key generation~\cite{kokoris2020asynchronous, gurkan2021aggregatable}, randomness beacons~\cite{syta2017scalable, abraham2023bingo, bhat2021randpiper}, federated learning~\cite{rathee2023elsa}, and distributed storage~\cite{dauterman2022waldo}.
It is also used to prevent malicious maximal extractable value attacks in blockchains~\cite{daian2020flash,qin2022quantifying}.
In such systems, performance and scalability are essential, since VSS protocols are often executed repeatedly and lie on the critical path of higher-level distributed applications.

In asynchronous environments, where message delays are unbounded and no timing assumptions can be made, asynchronous verifiable and complete secret sharing (AVSS and ACSS) provides the sharing layer needed by robust higher-level protocols and typically tolerates the optimal Byzantine bound $t < n/3$~\cite{hirt2021multi}.
Despite its importance, the practical behavior of modern one-shot AVSS and ACSS protocols remains poorly understood.
Recent constructions~\cite{abraham2023bingo,alhaddad2021high,yan2022shanrang,vassantlal2022cobra,yurek2022hbacss,yurek2023long,shoup2024lightweight, das2025verifiable,Xing2025GossamerLightweightAnd} combine different communication substrates, verification mechanisms, batching methods, and fault-handling paths.
Their theoretical guarantees are strong, but their concrete costs depend on how these mechanisms interact.
Practitioners therefore lack a principled basis for choosing a protocol under concrete latency, throughput, bandwidth, CPU, setup, and fault-assumption constraints.

Understanding asynchronous verifiable and complete sharing requires moving beyond protocol-centric analysis toward a design-space perspective.
In this paper, rather than viewing protocols as monolithic constructions, we decompose them into a set of fundamental design dimensions that capture how they disseminate information, ensure consistency, handle faults, and amortize cryptographic costs.
We show that design choices such as the use of reliable broadcast versus information dispersal, optimistic versus robust-first execution paths, and batching versus single-secret sharing have a direct and measurable impact on latency, throughput, systems costs (CPU, bandwidth, and message count), and scalability, and can make two protocols with similar asymptotic guarantees behave very differently.

To support this perspective, we present, to our knowledge, the first \textit{design-space analysis} of modern one-shot AVSS and ACSS constructions, connecting recurring protocol design patterns to implementation costs.
We implement a representative set of recent protocols within a common codebase and evaluate them under both optimistic (fault-free) and faulty-dealer stress conditions.
This controlled setting enables apples-to-apples comparisons and helps attribute observed costs to recurring protocol patterns rather than implementation artifacts.

Our empirical results reveal that no single protocol dominates across all metrics.
Instead, the selected protocols naturally cluster into distinct design archetypes, each corresponding to a different point in the design space.
Optimistic protocols achieve excellent performance in fault-free executions but degrade sharply under faulty-dealer stress.
Batch-oriented protocols scale efficiently but introduce latency overheads for small workloads.
High-resilience designs provide strong security guarantees at the cost of substantial communication and cryptographic overhead.
These outcomes are not incidental but stem from a small number of recurring mechanisms.

Building on these insights, we derive practical selection guidelines for AVSS and ACSS deployments.
They explain when to prioritize lightweight verification, predictable fault handling, batch amortization, packed communication, or high-threshold guarantees.
More broadly, our findings identify the sources of performance bottlenecks and inform the design of distributed cryptographic systems that balance strong security guarantees with practical efficiency.
In summary, we make the following contributions.
\begin{itemize}[left=0pt, itemsep=0pt,topsep=0pt,parsep=1pt]
      \item \textbf{Corpus-backed design-space analysis:}
            We build a curated corpus of modern VSS, AVSS, and ACSS work and derive the design dimensions that explain the main systems costs of asynchronous secret sharing.

      \item \textbf{Design-driven comparison of representative protocols:}
            We map seven representative protocols onto this design space and compare their assumptions, guarantees, adversary models, completion mechanisms, and asymptotic costs using a common vocabulary.

      \item \textbf{Common implementation and measurement framework:}
            We implement the selected protocols in a shared Rust framework and evaluate them under controlled honest-dealer and faulty-dealer stress executions, reporting performance metrics (latency and throughput) and cost metrics (messages, bandwidth, and CPU-time).

      \item \textbf{Design-to-cost findings and research agenda:}
            We show how recurring design patterns shape measured performance, derive deployment guidance, release reproducibility artifacts, and identify future research directions for recovery, batching, dissemination, verification, and hardware-assisted sharing.
\end{itemize}

The paper is organized as follows. Section~\ref{sec:related} reviews related work; Section~\ref{sec:sel} presents the VSS research landscape and selection methodology;
Section~\ref{sec:schemes} analyzes the seven selected one-shot sharing protocols; Section~\ref{sec:exp} describes the experimental framework; and Sections~\ref{sec:eval} through~\ref{sec:conclusion} report results, discuss implications, and conclude.

\section{Background and Related Work}\label{sec:related}

This section introduces the system model and terminology used throughout the paper, then reviews the most relevant related work.

\subsection{System model and notation}\label{sec:background}

We study one-shot sharing executions among $n$ parties $P_1,\dots,P_n$ and a designated dealer $D$ over authenticated point-to-point channels in a fully asynchronous network.
Messages may be delayed arbitrarily, but messages sent between honest parties are eventually delivered.
A party is \emph{honest} if it follows the protocol, and the adversary controls at most $t$ Byzantine parties that may collude, equivocate, omit messages, or act as a Byzantine dealer.
Some protocols also expose a privacy threshold $p$.
The threshold $t$ bounds Byzantine fault-tolerance, while $p$ bounds the coalition size against which the privacy of secrets is required.

A sharing execution outputs a \emph{sharing object}, which may be a single share, a vector of batched shares, a packed sharing, or an auxiliary transcript sufficient to use the share later.
Protocols typically have a \emph{sharing} phase, where $D$ distributes this object, and a \emph{reconstruction} phase, where parties later reconstruct the shared value.
Our benchmark isolates the one-shot sharing phase and invokes reconstruction only when its guarantees explain validity or recovery.

Because the selected papers use related but non-identical property names, we use the following comparison vocabulary.
\underline{Correctness} or \underline{honest-dealer validity} means that if $D$ is honest and inputs $s$ (or a vector of secrets), every honest party that outputs obtains a valid share of $s$, and sufficiently many valid shares reconstruct $s$.
\underline{Agreement} is a consistency property that means that honest parties do not accept incompatible sharing objects for the same execution, and honest outputs are valid for one canonical object reconstructed to the same value $\tilde{s}$, with $\tilde{s}=s$ when $D$ is honest.
\underline{Privacy} means that when $D$ is honest, the adversary's view during sharing reveals nothing about $s$ beyond corrupted shares and public metadata, in the computational or statistical sense specified by the protocol.
\underline{Termination} means that if $D$ is honest and honest messages are eventually delivered, all honest parties eventually finish sharing with a valid output.
\underline{Sharing completion} means that once any honest party accepts a usable sharing output, every honest party eventually obtains a compatible usable output for the same sharing object, unless the protocol specifies a common default, abort, or attributable failure outcome.
This is the property we compare across AVSS and ACSS variants.
It is weaker than requiring all protocols to realize the same ACSS functionality, but stronger than ordinary VSS verifiability.
Some papers call this condition agreement, completeness, or totality.
In this paper, \emph{agreement} denotes consistency of accepted outputs, while \emph{sharing completion} denotes propagation of an accepted sharing outcome to all honest parties.

\subsection{Related work}

Prior work relevant to this study falls into four categories, namely individual AVSS/ACSS evaluations, systems benchmarking, protocol-engineering studies, and survey or SoK-style taxonomies.
Recent AVSS and ACSS papers usually evaluate their own construction under protocol-specific codebases, platforms, parameters, and metrics.
\avss reports dealing and verification micro-benchmarks~\cite{das2025verifiable}; \hbacss evaluates scalability and fault-handling overhead~\cite{yurek2022hbacss}; \honey measures the broader dynamic proactive system built around its high-threshold sharing component~\cite{yurek2023long}; \goss reports WAN experiments for lightweight batched ACSS~\cite{Xing2025GossamerLightweightAnd}; and \light and \bingo mainly emphasize asymptotic analysis or proof-of-concept measurements~\cite{shoup2024lightweight,abraham2023bingo}.
These results are valuable, but they do not sufficiently normalize implementations and workloads to attribute performance differences to AVSS/ACSS design choices rather than the experimental context.

Distributed-systems research provides useful benchmarking and protocol-engineering precedents.
BlockBench compares permissioned blockchain platforms through a shared harness, common workloads, and component-level metrics such as throughput, latency, scalability, and fault tolerance~\cite{dinh2017blockbench}.
BFT studies similarly use implementation-driven evaluation to expose architectural costs, including robust asynchronous consensus in HoneyBadgerBFT~\cite{miller2016honey} and trusted-component trade-offs in BFT consensus~\cite{gupta2023dissecting}.
These works motivate our use of common measurement, faulty-dealer stress scenarios, and design-to-cost interpretation.
However, they target consensus, replicated state-machine execution, or blockchain platforms, not one-shot verifiable and complete sharing.
They therefore do not answer which AVSS/ACSS mechanisms--reliable broadcast, information dispersal, polynomial commitments, batching, packing, or recovery--drive bandwidth, CPU time, message count, and latency.

Survey and SoK-style studies show how to organize divergent protocol families through explicit design dimensions.
Bano et al. systematize blockchain consensus through security, performance, and design properties~\cite{bano2017consensus}, while Chandramouli et al.~\cite{chandramouli2022survey} survey perfectly secure VSS with emphasis on pre-2020 theoretical constructions.
Classical secret-sharing performance studies compare Shamir sharing~\cite{shamir1979share}, Krawczyk-style information dispersal~\cite{krawczyk1993secret,rabin1989efficient}, and storage-oriented deployments~\cite{li2015cdstore,bessani2013depsky,miranda2017experimental,nirmala2012comparative,abdallah2015secret,von2018performance}, but generally do not provide asynchronous completion, Byzantine-dealer verifiability, or Byzantine fault handling.
The gap is therefore specific.
No prior work connects modern AVSS/ACSS design choices to measured systems costs under common implementation and faulty-dealer stress scenarios.

\section{One-shot AVSS/ACSS design space and protocol selection}
\label{sec:method}
\label{sec:sel}

In this section, we define the scope of our study, construct a design-space sieve for recent one-shot asynchronous sharing protocols, and select representative protocols for theoretical analysis and empirical measurement.

\subsection{Scope and terminology}

We use \emph{verifiable secret sharing} (VSS) as the broad class of protocols in which parties can verify that their shares are consistent with a well-defined secret, even when the dealer or some participants are Byzantine.
\emph{Asynchronous VSS} (AVSS) provides this guarantee without timing assumptions on message delivery.
Minimal AVSS definitions may only require verifiable reconstructability, but recent practical AVSS protocols usually add an \emph{output propagation} guarantee because applications such as MPC preprocessing, asynchronous DKG, and randomness generation need all honest parties to obtain usable shares.
\emph{Asynchronous complete secret sharing} (ACSS) names this requirement explicitly.
Once one honest party outputs, all honest parties eventually obtain compatible shares of the same sharing object.
Some papers retain the AVSS name while providing the sharing-completion property defined in Section~\ref{sec:background}; therefore, we keep the terminology used by the original papers and state the corresponding guarantee explicitly.
Our study focuses on the one-shot sharing interface in which a dealer distributes one or more secrets, honest parties obtain verifiable and mutually consistent shares, and recovery is invoked only to complete that sharing execution.
We compare AVSS and ACSS through this interface even when a paper also uses the sharing protocol as a building block for longer-lived services.

\subsection{Research landscape and filtering}

We conducted a structured literature search covering the last decade.
The search combined keyword queries on Google Scholar and the IACR ePrint archive with a manual review of major security, systems, and cryptography venues, including ACM CCS, USENIX Security, IEEE S\&P, NDSS, CRYPTO, EUROCRYPT, and ASIACRYPT.
We retained papers that introduce new VSS, AVSS, or ACSS protocol constructions, as well as broader asynchronous services that specify a one-shot verifiable or complete sharing component.
We then identified the subset whose asynchronous setting includes explicit verifiability together with sharing completion or an equivalent first-output propagation guarantee.
We excluded protocols whose primary contribution lies outside this setting, such as visual or quantum secret sharing, homomorphic or function secret sharing, and schemes that do not address asynchronous Byzantine behavior.
We also filtered out works that only instantiate secret sharing inside a larger application, optimize a local algebraic primitive, or study synchrony models outside the asynchronous Byzantine setting targeted here.

\begin{figure}
    \centering
    \includegraphics[width=\linewidth]{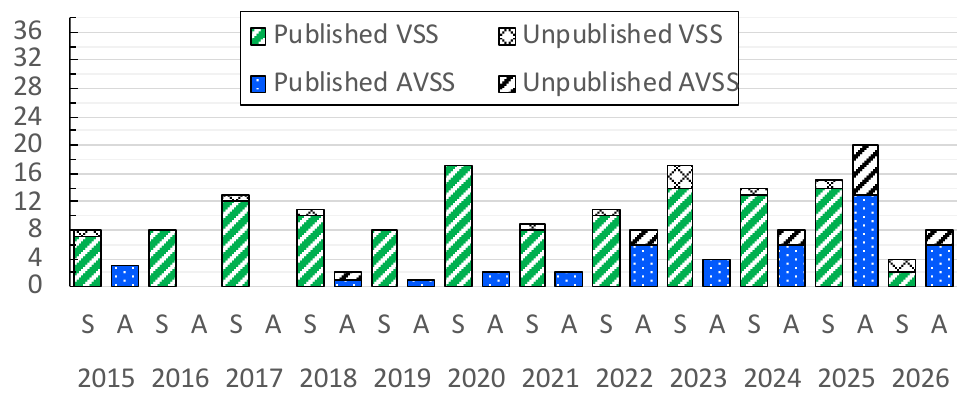}
    \caption{Research landscape of VSS (S) and asynchronous sharing protocols in scope (A).}
    \label{fig:selection}
\end{figure}

This process yielded a curated corpus of 193 VSS-related papers under different synchrony assumptions.
We release the corpus, coding sheet, and selection metadata\footnote{https://doi.org/10.5281/zenodo.21417706}, including source papers, inclusion and exclusion decisions, venue tags, and the mapping to the representative protocols evaluated here.
As shown in Figure~\ref{fig:selection}, the literature has expanded rapidly over the last decade and has become increasingly heterogeneous.
Only 35 published papers in the corpus address or contain an AVSS or ACSS sharing interface in this scope, and nearly three quarters of those appeared since 2022.
This recent acceleration makes exhaustive implementation impractical and, more importantly, methodologically weak.
Many constructions differ in secondary details while sharing the same dominant mechanisms.
We therefore use the corpus to identify the recurring design dimensions of the field, then evaluate representative protocols that cover those dimensions.

\subsection{A design-space sieve}

The sieve is defined before selecting concrete protocols.
Because the corpus is restricted to protocols that provide one-shot AVSS or ACSS semantics with sharing completion, completion is not a binary inclusion axis at this stage.
Instead, it asks how completion is obtained, verified, recovered, and amortized.
For each candidate protocol, we first record its security envelope, namely its corruption model and privacy threshold, because these assumptions determine whether a protocol is applicable to a given system.
Within that envelope, five design elements explain the main performance trade-offs observed in our evaluation.

\paragraph{Communication model}
The communication model determines how shares, commitments, and completion evidence become available to honest parties.
Modern one-shot protocols range from direct dealer-to-party secure message distribution, to reliable-broadcast transcripts, to encrypt-then-disperse designs based on AVID, to peer exchange over algebraic sharing objects.
Some constructions keep completion evidence private unless a complaint occurs, whereas others publish enough transcript data for all parties to verify the sharing outcome.
This dimension controls message volume, synchronization points, and scalability as $n$ grows.

\paragraph{Cryptographic verification model}
The verification model determines how a party decides that a share is safe to use.
The corpus spans hash and Merkle evidence, hash-derived degree checks for bivariate shares, polynomial or vector commitments, public-key encryption, and zero-knowledge proofs.
These choices also encode setup assumptions, such as PKI, trusted setup, common reference strings, or no setup.
They determine whether a protocol is primarily network-bound, CPU-bound, or dominated by proof generation and verification.

\paragraph{Recovery strategy}
Sharing completion is most expensive when the dealer is faulty, slow, or equivocating.
Candidate protocols handle this case through complaint-driven recovery, transcript broadcast, AVID disclosure and interpolation, original-evaluation forwarding, interpolation from verified peer data, or abort-and-retry behavior.
The recovery strategy determines whether a protocol merely detects an invalid sharing or completes despite dealer faults.
It is therefore the main explanation for the gap between honest-dealer and faulty-dealer stress executions.

\paragraph{Batching and packing}
Modern ACSS workloads rarely consist of a single isolated secret.
Protocols can share independent secrets in a batch, pack several secrets into one algebraic object, or use information dispersal to make large payloads cheap after a small secret has been shared.
This dimension distinguishes protocols with cryptographic batch amortization, protocols that pack several secrets into one algebraic sharing instance, vectorized designs that group several independent sharing polynomials into one multi-secret execution, and protocols that are naturally evaluated as single-secret designs.
This dimension drives when throughput improves with larger workloads and when batching only increases latency.

\paragraph{Execution model}
Finally, protocols differ in when they pay the cost of robustness.
Optimistic protocols keep the fault-free path short and defer expensive work to complaints or recovery.
Robust-first protocols pay more in every execution to reduce variance when faults occur.
Intermediate designs keep a recognizable happy path but still pay for output-propagation evidence, recovery metadata, or proof-carrying dissemination in every execution.
This dimension directly matches our optimistic and faulty-dealer stress experiments.

\subsection{Selection criteria}

We select concrete protocols only after defining the sieve.
A protocol is included in the benchmark if it satisfies four criteria.
First, it exposes a one-shot asynchronous sharing interface with AVSS or ACSS semantics under Byzantine faults.
Second, it appeared in the 2021--2026 wave of modern asynchronous sharing protocols and is not merely a minor variant of another selected construction.
Third, the paper specifies the protocol sufficiently to support a faithful implementation in our common framework.
Fourth, the final set is chosen to maximize design diversity across the sieve rather than exhaustively cover the literature.
When a paper develops a broader service, we evaluate only the one-shot sharing component that satisfies these criteria.

\subsection{Representative design points}

Applying these criteria yields seven representative protocols: \haven, \hbacss, \avss, \bingo, \goss, \light, and \honey.
They are representative design points, not an exhaustive SoK of every protocol in the corpus.

\light~\cite{shoup2024lightweight} is a hash-based AVSS construction, using secure message distribution, a random beacon, compact broadcast, and one-sided voting.
It minimizes cryptographic cost and amortizes well in the fault-free case, but its complaint path is expected to create a large performance gap between honest-dealer and faulty-dealer stress executions.
\hbacss~\cite{yurek2022hbacss} represents the implemented hbACSS0-style point in the hbACSS family: batch-oriented ACSS based on AVID dispersal, batch polynomial commitments, and key-disclosure share recovery.
\avss~\cite{das2025verifiable} represents a compact, publicly verifiable AVSS design with transcript-based completion and dual-threshold support; its multi-secret measurements use the vectorized interface defined in Section~\ref{sec:exp}.
\bingo~\cite{abraham2023bingo} represents adaptive security and packed AVSS/ACSS through bivariate polynomials.
\goss~\cite{Xing2025GossamerLightweightAnd} represents lightweight batched ACSS with linear amortized per-secret communication based on original-evaluation propagation and bivariate degree checking.
\haven~\cite{alhaddad2021high} represents high-threshold, dual-threshold AVSS with no-setup robustness, while \honey~\cite{yurek2023long} represents encrypted high-threshold complete sharing with public-key verification.
Table~\ref{tab:ss_comparison} applies the sieve to the selected protocols and exposes their design differences.

\begin{table*}[t]
        \caption{Selected one-shot AVSS/ACSS design points.}
        \label{tab:ss_comparison}
        \centering
        \begingroup
        \setlength{\tabcolsep}{2pt}
        \footnotesize
        \begin{tabular}{@{}
                        P{0.12\textwidth}
                        P{0.14\textwidth}
                        P{0.17\textwidth}
                        P{0.12\textwidth}
                        P{0.21\textwidth}
                        P{0.13\textwidth}
                        P{0.055\textwidth}
                        @{}}
                \toprule
                \textbf{Protocol}
                 & \textbf{\makecell[c]{Adversary          \\model}}
                 & \textbf{\makecell[c]{Communication      \\model}}
                 & \textbf{\makecell[c]{Verification       \\model}}
                 & \textbf{\makecell[c]{Fault handling /   \\completion}}
                 & \textbf{\makecell[c]{Batching/packing}}
                 & \textbf{\makecell[c]{Happy              \\Path ?}}                                \\
                \midrule
                \makecell[c]{\haven~\cite{alhaddad2021high}\\{\emph{\scriptsize AVSS}}}
                 & Adaptive; Byzantine
                 & RBC-heavy dissemination
                 & Vector commitments
                 & Output propagation
                 & No native batching
                 & \ding{56}                               \\
                \midrule
                \makecell[c]{\hbacss~\cite{yurek2022hbacss}\\{\emph{\scriptsize ACSS}}}
                 & Static; Byzantine
                 & Encrypt-then-disperse with AVID
                 & PC-based validation
                 & Share recovery
                 & Batched sharing
                 & \ding{51}                               \\
                \midrule
                \makecell[c]{\bingo~\cite{abraham2023bingo}\\{\emph{\scriptsize AVSS/ACSS}}}
                 & Adaptive; Byzantine
                 & Peer exchange over bivariate shares
                 & PC checks
                 & Share recovery
                 & Packs $t+1$ secrets
                 & \ding{56}                               \\
                \midrule
                \makecell[c]{\honey~\cite{yurek2023long}\\{\emph{\scriptsize ACSS}}}
                 & Static; PPT Byzantine
                 & RBC and public-key encryption
                 & PC and ZKP
                 & Transcript/default
                 & No native batching
                 & \ding{56}                               \\
                \midrule
                \makecell[c]{\light~\cite{shoup2024lightweight}\\{\emph{\scriptsize AVSS}}}
                 & Static; Byzantine
                 & Dealer-to-party SMD
                 & Hashes and Merkle proofs
                 & Share recovery
                 & Batched sharing
                 & \ding{51}                               \\
                \midrule
                \makecell[c]{\avss~\cite{das2025verifiable}\\{\emph{\scriptsize ACSS}}}
                 & Static; PPT Byzantine
                 & Dealer-to-party plus RBC transcript
                 & PC checks
                 & Transcript completion
                 & Vectorized batching
                 & \ding{56}                               \\
                \midrule
                \makecell[c]{\goss~\cite{Xing2025GossamerLightweightAnd}\\{\emph{\scriptsize ACSS}}}
                 & Static; PPT Byzantine
                 & SMD and original-evaluation propagation
                 & Hash-based degree checks
                 & Share recovery
                 & Batched ACSS
                 & \ding{51}                               \\
                \bottomrule
        \end{tabular}
        \endgroup
        \vspace{1mm}
        \begin{minipage}{0.98\textwidth}
                \footnotesize
                RBC, AVID, PC, ZKP, and SMD denote reliable broadcast, asynchronous verifiable information dispersal, polynomial commitments, zero-knowledge proofs, and secure message distribution, respectively.
                In the fault-handling column, share recovery denotes sharing-phase repair for missing valid shares, transcript completion denotes completion from an agreed transcript, transcript/default denotes a transcript-based path where invalid transcripts lead to abort or default output, and output propagation denotes completion from the agreement property without a separate repair branch.
                Vectorized batching denotes a multi-secret execution built from independent sharing polynomials whose messages are grouped, without packing several secrets into one algebraic object.
                \emph{Adversary terminology.}
                In the adversary-model column, static/adaptive refers to corruption timing, PPT refers to computational boundedness.
        \end{minipage}
        \vspace{-2mm}
\end{table*}

Several protocols in the corpus are not implemented in our current benchmark but are important for interpreting the design space.
COBRA~\cite{vassantlal2022cobra} and Shanrang~\cite{yan2022shanrang} refine related asynchronous agreement and sharing services, but their recovery, deployment, or synchrony assumptions differ from the fully asynchronous sharing-completion setting that our experiments isolate.
Some other in-scope constructions are not implemented because they refine design points already covered by the benchmark rather than adding an orthogonal axis.
Ji--Li--Song ACSS~\cite{Ji2024LinearCommunicationAsynchronous} advances the linear-communication ACSS frontier, which is already represented experimentally by \goss.
\haven++~\cite{Alhaddad2024HavenBatchedAnd} extends the \haven line with batching and packing, while our set already separates high-threshold sharing, batching, and packing through \haven, \hbacss, \bingo, and \goss.
We therefore use these works to position the frontier of the corpus, while keeping the implemented set focused on design diversity under a common measurement framework.
Our experimental measurements explain the behavior of implemented representative design points, while the sieve provides a common vocabulary for positioning newer constructions.

The design-space analysis leads to five evaluation questions that we answer in Section~\ref{sec:eval} by relating each observed performance trend back to the design dimensions above.
\begin{enumerate}[left=0pt, itemsep=0pt,topsep=0pt,parsep=1pt]
    \item How large is the gap between optimistic and faulty-dealer stress executions?
    \item Which communication models dominate latency and throughput as the number of parties grows?
    \item Which cryptographic verification mechanisms become bottlenecks in practice?
    \item When does batching or packing amortize fixed costs, and when does it only increase latency?
    \item Which fault-handling strategies best preserve progress under faulty-dealer stress?
\end{enumerate}

\section{Selected one-shot sharing protocols}
\label{sec:schemes}

This section interprets the seven selected protocols as design points in the sieve rather than as isolated constructions.
Tables~\ref{tab:ss_comparison} and~\ref{tab:ss_comparison2} summarize the security envelopes, architectural choices, and asymptotic costs, while Table~\ref{tab:costly-components} identifies the dominant design-level cost drivers.
For each protocol, we separate its research contribution, design pattern, operational sharing sketch, and performance implication.

\begin{table*}[t]
	\caption{Normalized asymptotic cost of selected one-shot sharing interfaces.}
	\label{tab:ss_comparison2}
	\centering
	\begingroup
	\setlength{\tabcolsep}{1.5pt}
	\scriptsize
	\begin{tabular}{@{}
			P{0.10\textwidth}
			P{0.11\textwidth}
			P{0.08\textwidth}
			P{0.07\textwidth}
			P{0.20\textwidth}
			P{0.15\textwidth}
			P{0.07\textwidth}
			P{0.14\textwidth}
			@{}}
		\toprule
		\textbf{Scheme}
		 & \textbf{\makecell[c]{Shared                               \\object}}
		 & \textbf{\makecell[c]{Thresholds                           \\$(t,p)$}}
		 & \textbf{\makecell[c]{Total                                \\msg.}}
		 & \textbf{\makecell[c]{Total comm.                          \\H/F}}
		 & \textbf{\makecell[c]{Amortized                            \\comm.\\per secret}}
		 & \textbf{\makecell[c]{Steps                                \\H/F}}
		 & \textbf{\makecell[c]{Setup/                               \\assumptions}} \\
		\midrule
		\haven\cite{alhaddad2021high}
		 & \makecell[c]{1 random secret;                             \\large via IDA}
		 & \makecell[c]{$t<n/3$                                      \\$p<n-t$}
		 & $\bigO(n^2)$
		 & \makecell[c]{short: $\bigO(\kappa n^2\log n)$             \\large: $\bigO(n|s|)$}
		 & same as total
		 & 3 / 3
		 & \makecell[c]{No PKI/setup;                                \\DL, ROM} \\
		\midrule
		\makecell[c]{\hbacss~\cite{yurek2022hbacss}\\{\scriptsize hbACSS0 rec.}}
		 & \makecell[c]{$b$ secrets                                  \\batched}
		 & \makecell[c]{$t<n/3$                                      \\$p=t$}
		 & $\bigO(n^2)$
		 & \makecell[c]{$\bigO(\kappa b n\log n)$                    \\/\ $\bigO(\kappa b n^2\log n)$}
		 & \makecell[c]{$\bigO(\kappa n\log n)$                      \\/\ $\bigO(\kappa n^2\log n)$}
		 & 4 / 6
		 & \makecell[c]{PKI, no SRS;                                 \\DL, ROM} \\
		\midrule
		\bingo\cite{abraham2023bingo}
		 & \makecell[c]{$t+1$ secrets                                \\packed}
		 & \makecell[c]{$t<n/3$                                      \\$p=t$}
		 & $\bigO(n^2)$
		 & $\bigO(\lambda n^2)$
		 & $\bigO(\lambda n)$
		 & 6 / 6
		 & \makecell[c]{PKI, powers-of-tau;                          \\q-SDH, AGM} \\
		\midrule
		\honey\cite{yurek2023long}
		 & 1 secret
		 & \makecell[c]{$t<n/3$                                      \\$p<n-t$}
		 & $\bigO(n^2)$
		 & $\bigO(\kappa n^2)$
		 & $\bigO(\kappa n^2)$
		 & 3 / 3
		 & \makecell[c]{PKI, Pedersen/NIZK;                          \\DL, Paillier, ROM} \\
		\midrule
		\light\cite{shoup2024lightweight}
		 & \makecell[c]{$b$ secrets                                  \\batched}
		 & \makecell[c]{$t<n/3$                                      \\$p=t$}
		 & $\bigO(n^2)$
		 & \makecell[c]{$\bigO(\kappa b n\log n+\kappa n^2\log n)$   \\/\ $\bigO(\kappa b n^2\log n+\kappa n^3\log n)$}
		 & \makecell[c]{$\bigO(\kappa n\log n)$                      \\/\ $\bigO(\kappa n^2\log n)$}
		 & 9 / 11
		 & \makecell[c]{beacon, hash-based}                          \\
		\midrule
		\avss\cite{das2025verifiable}
		 & \makecell[c]{$b$ secrets                                  \\vectorized}
		 & \makecell[c]{$t<n/3$                                      \\$p<n-t$}
		 & $\bigO(n^2)$
		 & $\bigO(\kappa b n^2)$
		 & $\bigO(\kappa n^2)$
		 & 5 / 5
		 & \makecell[c]{PKI;                                         \\DL, sig./VE} \\
		\midrule
		\goss\cite{Xing2025GossamerLightweightAnd}
		 & \makecell[c]{$b$ secrets                                  \\batched}
		 & \makecell[c]{$t<n/3$                                      \\$p=t$}
		 & $\bigO(n^3)$
		 & \makecell[c]{$\bigO(\kappa b n+\kappa n^3\log n)$       }
		 & $\bigO(\kappa n+\kappa n^3\log n/b)$
		 & 3 / 6
		 & \makecell[c]{No CRS;                                      \\hash/ROM} \\
		\bottomrule
	\end{tabular}
	\endgroup
	\vspace{1mm}
	\begin{minipage}{0.98\textwidth}
		\scriptsize
		All costs are per secret sharing instance.
		Messages are point-to-point protocol sends by honest parties, total communication is whole-network traffic, and amortized communication divides it by the number of output secrets.
		H/F denotes honest-dealer execution versus faulty-dealer stress execution.
		Step counts are asynchronous communication steps, inferred from the protocol schedule when not stated explicitly. Bracha-style reliable broadcast counts as three steps.
		For batched protocols, $b$ is the batch size; \bingo packs $t+1$ secrets.
		The \hbacss row reports the hbACSS0-style recovery path used in our implementation; hbACSS1/2 use different recovery mechanisms and have different faulty-path costs.
		The \light row uses the basic beacon-based construction reported in the original paper~\cite{shoup2024lightweight}; the ROM three-move optimization has a different round schedule and is not the measured variant.
	\end{minipage}
\end{table*}

\begin{table}[ht]
    \caption{Dominant happy-path and recovery-path cost drivers.}
    \label{tab:costly-components}
    \centering
    \setlength{\tabcolsep}{1.5pt}
    \footnotesize{
        \begin{tabular}{
            @{}
            P{0.09\textwidth}
            >{\raggedright\arraybackslash}p{0.2\textwidth}
            >{\raggedright\arraybackslash}p{0.18 \textwidth}
            @{}
            }
            \toprule
            \textbf{Scheme}
             & \textbf{Happy-Path Driver}
             & \textbf{Recovery-Path Driver}                                                       \\
            \midrule

            \haven
             & Polynomial/vector commitment dissemination; Bracha-style RBC on the root commitment
             & No recovery path; completion/reconstruction use commitment-checked evaluations      \\
            \midrule

            \hbacss
             & AVID dispersal/retrieval and polynomial-commitment proof generation/verification
             & Implication-driven AVID retrieval and share recovery                                \\
            \midrule

            \bingo
             & Bivariate KZG commitments, partial evaluations, and commitment RBC
             & Proof interpolation and row/column share recovery                                   \\
            \midrule

            \honey
             & Paillier/Pedersen ciphertext-and-proof bundle broadcast by RBC
             & No recovery path, RBC carries all encrypted shares/proofs                           \\
            \midrule

            \light
             & SMD/Merkle dissemination, hash checks, and one-sided voting
             & Unhappy-path SMD forwarding and robust interpolation                                \\
            \midrule

            \avss
             & Vectorized dealer-to-party sharing, signed ACKs, and transcript RBC
             & Per-secret transcript completion with batched verification of missing-party openings                  \\
            \midrule

            \goss
             & SMD original-evaluation propagation, BP-DC checks, and one-sided voting
             & Dealer-disclosure evidence, SMD forwarding, and robust interpolation                \\

            \bottomrule
        \end{tabular}
    }
\end{table}

\subsection{\haven ~- High-Threshold AVSS}

\textbf{\emph{Research contribution.}}
\haven's core contribution is to combine high-threshold privacy, optimal asynchronous Byzantine resilience, and no trusted setup in a construction with lower communication than earlier high-threshold AVSS designs.

\textbf{\emph{Design pattern.}}
\haven~\cite{alhaddad2021high} uses a dual-threshold AVSS structure in which a two-layer polynomial-sharing construction, commitments, and reliable broadcast preserve agreement and privacy even when the privacy threshold $p$ exceeds the Byzantine threshold $t$.
For large secrets, it combines AVSS with symmetric encryption and information dispersal to reduce payload overhead.

\textbf{\emph{Sharing in a nutshell.}}
The dealer samples a degree-$p$ recovery polynomial $R$ with $R(0)=s$ and, for each party $P_i$, a degree-$t$ share polynomial $S_i$ constrained by $S_i(i)=R(i)$.
It commits to $R$, the $S_i$, and consistency tests, then runs a Bracha-style send/echo/ready pattern in which parties verify the root commitment, forward one certified point on each recipient's share polynomial, and converge on a common root.
A party completes by interpolating its own $S_i$ from valid echo data associated with the agreed root; reconstruction later reveals certified values $S_i(i)$, and any $p+1$ valid values interpolate $R(0)$.
Thus, \haven does not expose a separate sharing-phase repair branch; completion follows from the reliable-broadcast-style propagation of the agreed root and checked evaluations.

\textbf{\emph{Performance implication.}}
\haven pays for robustness upfront.
Broadcast-heavy dissemination and verification of many related polynomial commitments create a high baseline cost even with an honest dealer, yielding stable behavior across the honest-dealer and faulty-dealer stress scenarios but limited scalability compared with protocols that defer recovery work.

\subsection{\hbacss ~- Batch-Oriented ACSS}

\textbf{\emph{Research contribution.}}
\hbacss's core contribution is to make ACSS practical for asynchronous MPC-style workloads by combining AVID-style dissemination with batch polynomial commitments, so completion under faults becomes predictable rather than an exceptional slow path.

\textbf{\emph{Design pattern.}}
\hbacss~\cite{yurek2022hbacss} follows an encrypt-then-disperse pattern.
The dealer encrypts recipient-specific shares and proofs, disperses the ciphertexts through AVID, and uses agreement to ensure that enough parties received valid data.

\textbf{\emph{Sharing in a nutshell.}}
The dealer samples a batch of degree-$t$ sharing polynomials, reliably broadcasts the polynomial commitments, and disperses encrypted recipient payloads containing shares and batched evaluation proofs.
Each party retrieves and decrypts its payload, verifies the batched proofs against the broadcast commitments, and enters an OK/READY agreement path before outputting its shares.
In the implemented hbACSS0-style recovery path, a recipient that detects an invalid payload implicates the dealer by disclosing the disputed decryption key; parties that already hold valid shares then disclose their keys, and missing recipients retrieve, decrypt, verify, and interpolate at least $t+1$ valid payloads to complete.

\textbf{\emph{Performance implication.}}
The protocol pays a fixed cryptographic and dispersal overhead that is expensive for a single secret but amortizes over large batches.
Under faulty-dealer stress, disclosure and interpolation are triggered instead of an abort, yielding higher single-secret latency but stable large-batch throughput and a small gap relative to the honest-dealer scenario.

\subsection{\bingo ~- Adaptively Secure Packed AVSS}

\textbf{\emph{Research contribution.}}
\bingo's core contribution is to bring adaptive security to efficient packed AVSS/ACSS, showing that stronger corruption timing can be supported while amortizing several secrets within one bivariate sharing instance.

\textbf{\emph{Design pattern.}}
\bingo~\cite{abraham2023bingo} encodes $t+1$ secrets in a bivariate polynomial and uses KZG-based commitments so that parties can recover missing rows or columns from verified peer data.
Recovery is embedded in the row/column structure rather than isolated as a separate complaint protocol.

\textbf{\emph{Sharing in a nutshell.}}
The dealer embeds the packed secrets in a bivariate polynomial $\phi(X,Y)$, reliably broadcasts a bivariate commitment, and privately sends party $P_i$ its row polynomial $\phi(X,\omega_i)$ together with the hiding row.
Parties that receive valid rows evaluate them for other parties, enabling recipients to interpolate column evidence and then reconstruct missing rows from $2t+1$ verified column messages.
After enough parties have row and column material, they exchange done messages; reconstruction of one packed secret requires parties to open the corresponding row evaluation, and $t+1$ valid openings allow the relevant polynomial to be interpolated and evaluated at $Y=0$.

\textbf{\emph{Performance implication.}}
Packing reduces communication per secret, while adaptivity and bivariate verification introduce heavier cryptographic and peer-exchange costs.
This pattern favors \bingo when the workload matches its packing granularity and faults occur, but proof verification and peer exchange remain throughput limits under high concurrency.

\subsection{\honey ~- Encrypted High-Threshold Complete Sharing}

\textbf{\emph{Research contribution.}}
\honey's relevant contribution for this study is an encrypted high-threshold complete-sharing layer that combines dual-threshold privacy with publicly checkable consistency evidence.

\textbf{\emph{Design pattern.}}
\honey~\cite{yurek2023long} combines ACSS-style sharing with reliable broadcast, public-key encryption, Pedersen commitments, and zero-knowledge proofs.
In our benchmark, we isolate this one-shot sharing path and measure the cost of obtaining high-threshold privacy and completion through encrypted shares and proof-carrying broadcasts.

\textbf{\emph{Sharing in a nutshell.}}
For a fresh high-threshold sharing, the dealer samples sharing and blinding polynomials, reliably broadcasts encrypted evaluations, Pedersen commitments, and zero-knowledge proofs tying each ciphertext to its commitment.
Recipients verify the proofs and perform the degree check, decrypt their own share and blinding share, and keep public commitments that later authenticate reconstruction or transfer.
Completion follows from agreement on the broadcast transcript and from local verification of the encrypted evaluations against the commitments.
If the transcript is invalid, parties abort/default rather than running a peer-based share-recovery procedure.
Reconstruction multicasts share and blinding-share openings, verifies them against the public commitments, and interpolates the secret from enough valid openings.

\textbf{\emph{Performance implication.}}
\honey pays for high-threshold privacy and public verifiability in every sharing execution.
Public-key operations, zero-knowledge proofs, and reliable broadcast dominate its cost, so its latency reflects the price of strong one-shot sharing guarantees rather than a batching-oriented design.

\subsection{\light ~- Optimistic Hash-Based AVSS}

\textbf{\emph{Research contribution.}}
\light's core contribution is to show that practical AVSS can avoid public-key-heavy machinery in the common path, using hash and symmetric primitives for lightweight, post-quantum-friendly batched sharing while isolating the cost of faults.

\textbf{\emph{Design pattern.}}
\light~\cite{shoup2024lightweight} denotes the basic construction from the original paper, which is based on secure message distribution, a random beacon, compact broadcast, one-sided voting, hash/Merkle evidence, symmetric primitives, and batched consistency checks.
We do not use the later ROM three-move optimization when reporting step counts or experiments.
It keeps the honest-dealer path lightweight and moves expensive work to complaints and recovery.

\textbf{\emph{Sharing in a nutshell.}}
The dealer distributes, through secure message distribution, each party's evaluations of the batched sharing polynomials together with blinding evaluations used for a random linear degree check.
After a random beacon challenge, the dealer reliably broadcasts response polynomials, parties locally check the induced linear identities, and happy parties enter one-sided voting and output immediately once the vote succeeds.
An unhappy party forwards a verifiable complaint; happy parties answer with assists and reports, and the unhappy party interpolates its missing share vector from enough valid assists.

\textbf{\emph{Performance implication.}}
This design creates the clearest gap between happy-path and faulty-dealer stress executions.
It gives excellent latency and throughput with an honest dealer, followed by a pronounced slowdown when recovery communication and synchronization are triggered.

\subsection{\avss ~- Transcript-Based Public Verifiability}

\textbf{\emph{Research contribution.}}
\avss's core contribution is to simplify AVSS around publicly checkable transcripts and acknowledgment evidence, reducing interactive complaint complexity while making completion evidence externally auditable.

\textbf{\emph{Design pattern.}}
\avss~\cite{das2025verifiable} centers completion on a compact transcript.
Parties acknowledge valid shares, and the dealer broadcasts a transcript containing the commitment, acknowledgments, and openings for parties that did not acknowledge.

\textbf{\emph{Sharing in a nutshell.}}
In the asynchronous protocol, the dealer commits to a degree-$2t$ sharing polynomial, sends each party its private share and opening proof, and waits for $2t+1$ signatures acknowledging the commitment.
It then reliably broadcasts a public transcript containing the commitment, the acknowledgments, the set of missing signers, and batched openings for those missing shares.
Each party verifies the transcript and obtains its share either from its private message or from the transcript. Reconstruction multicasts share/opening pairs, and $2t+1$ valid pairs interpolate the committed polynomial at zero.
The batched operation described in the original paper is a batched verification of openings for several party indices.
For multi-secret workloads, our evaluation uses the vectorized interface defined in Section~\ref{sec:exp}: independent per-secret polynomials with grouped messages.
This is parallel composition, not packing or cryptographic batch amortization.

\textbf{\emph{Performance implication.}}
The transcript replaces interactive complaints with public evidence, which favors simple and robust single-secret execution.
In our experiments, grouped messages amortize scheduling and transport overhead, but transcript work and commitment verification remain per secret.

\subsection{\goss ~- Lightweight Linear-Communication ACSS}

\textbf{\emph{Research contribution.}}
\goss's core contribution is to remove homomorphic commitments from linear-communication ACSS, using original-evaluation propagation and bivariate degree checking to preserve amortized linear communication in both happy and recovery executions for sufficiently large batches.

\textbf{\emph{Design pattern.}}
\goss~\cite{Xing2025GossamerLightweightAnd} combines secure message distribution, batched original-evaluation propagation, and bivariate-polynomial degree checking.
Unlike commitment-based linear-communication protocols, it keeps the core ACSS path hash- and symmetric-key-based.

\textbf{\emph{Sharing in a nutshell.}}
The dealer partitions the batch into groups of $t+1$ secrets, embeds each group in a $(t,t)$ bivariate polynomial, sends each recipient its data in original-evaluation form through secure message distribution, and broadcasts a lightweight verification script $v(x,y)=r\cdot A(x,y)+E(x,y)$.
Recipients derive the Fiat-Shamir challenge from the SMD roots, verify their local evaluations against the script, and enter one-sided voting before decoding their row into shares.
If a faulty evaluation is detected, the node forwards it as dealer evidence; recovery then propagates only original authenticated evaluations so parties interpolate columns, redistribute row points, and reconstruct the missing rows without homomorphic commitments.

\textbf{\emph{Performance implication.}}
\goss trades a larger fixed batch overhead for linear amortized per-secret communication.
For ACSS, total communication is $\bigO(\kappa b n+\kappa n^3\log n)$ in both happy and recovery executions, and the $\bigO(\kappa n)$ per-secret cost appears once $b \geq \bigO(n^2\log n)$.
The key trade-off is whether \goss's higher fixed cost is repaid in large-batch and faulty-dealer stress executions.

\section{Experimental environment}
\label{sec:exp}

Fairly comparing one-shot asynchronous verifiable and complete sharing protocols requires measuring both performance and concrete resource costs under the same workloads and fault scenarios.
We evaluate each selected protocol under controlled optimistic and faulty-dealer stress executions while varying the number of parties and the batch size.
For each completed sharing execution, we report completion latency, peak throughput, protocol messages, serialized bytes, and aggregate CPU-time across all parties.
For a single-secret execution, these costs are per secret; for a batched execution, they are reported per completed batch.

Protocols are evaluated under two scenarios.
In the optimistic scenario, the dealer is honest and no parties are Byzantine, capturing best-case latency and throughput, with the full protocol executing without complaints or recovery.
In the faulty-dealer stress scenario, the dealer is Byzantine and therefore consumes one corruption in the $t$-fault budget.
It sends incorrect or missing shares to up to $t-1$ parties while behaving honestly toward the others, and those additional faulty parties remain silent to delay execution.
We call this a stress scenario because it deliberately drives protocols away from their complaint-free path and exercises their completion, detection, and fault-handling mechanisms under a controlled faulty-dealer pattern.
It is therefore a meaningful faulty-dealer stress test, not a protocol-level worst case, and it does not exhaust all Byzantine schedules, equivocation patterns, or adaptive corruptions.

\subsection{Implementation and framework}
\label{sec:implementation}

We present our framework for implementing and evaluating one-shot AVSS and ACSS protocols; it comprises approximately 13,700 lines of Rust code~\cite{anonymous_2026_21417658}.

Each protocol is implemented as an independent module over a shared interface and infrastructure, enabling easy extension by registering new modules.
The framework provides reusable subprotocols, \ie, Secure Message Distribution~\cite{shoup2024lightweight}, Reliable Broadcast~\cite{cachin2005asynchronous}, Disperse-Retrieve~\cite{das2021asynchronous}, and One-Sided Vote~\cite{shoup2024lightweight}, to avoid code duplication.
Communication is built on Tokio's asynchronous runtime~\cite{tokio2024}, using typed TCP messages dispatched to dedicated workers, while cryptographic primitives rely on established libraries~\cite{e2evss2023, apss2022} shared across protocols for fair comparison.
The framework also includes a distributed evaluation environment that executes protocols on multiple physical machines over real TCP connections, without hardware acceleration or multithreading.
Experiments are deployed via Ansible-managed daemons and coordinated by a central Rust orchestrator driven by a JSON configuration, which assigns protocol instances, manages execution, and generates plots from collected logs.
For fairness, all protocols use the same network stack, serialization layer, process placement, logging infrastructure, and shared implementations of common subprotocols whenever the specifications permit it.
We instantiate each protocol with the parameters prescribed by its paper, set the fault threshold as $t=\lfloor n / 3 \rfloor$ unless the protocol requires otherwise.
We cross-checked each implementation against the corresponding protocol description and verified that it reproduced the reported asymptotic trends.
Message counts are measured at the protocol layer, byte counts include serialized protocol payloads, and CPU-time aggregates the computation performed by all protocol processes while excluding orchestration and logging.

\subsection{Experimental Setup}

All experiments were conducted on the Grid'5000 testbed~\cite{cappello2005grid} using 20 identical machines from the Nova cluster, each equipped with an Intel Xeon E5-2620 v4 (8 cores), 64\,GiB RAM, and 10\,Gbps Ethernet.
The number of machines was chosen empirically to avoid CPU saturation, providing a homogeneous environment in which observed performance differences primarily reflect protocol-induced costs under this controlled setup.
Our goal is to compare protocol-induced costs rather than characterize a particular geo-distributed deployment.
Grid'5000 isolates the effects of communication patterns, verification, batching, and recovery, while messages, bytes, CPU-time, and step structure indicate which designs are most sensitive to WAN delay or bandwidth constraints.

We measure completion latency by running 100 sequential sharing instances per protocol. Because the observed distributions have low dispersion, we report the average time from the dealer's start until all honest parties have produced an output.
For the same executions, we record the total protocol messages, serialized bytes, and aggregate CPU-time across all parties.
CPU-time is measured inside each Rust process with the Rust \texttt{cpu-time} crate, whose library interface is \texttt{cpu\_time}, and then summed across protocol processes.
To measure peak performance, we run controlled-load experiments at increasing input rates $r_i$ for 300\,s and, for each $r_i$, average the achieved output rate $o_i$ in secrets/s and the 99th-percentile completion latency $L_i$ over 10 runs.
The peak-throughput knee is the last stable point before saturation: the first index $i$ for which increasing the input rate to $r_{i+1}$ yields no material throughput gain, while at least doubling tail latency.
We report $o_i$ as peak throughput, measure latency and resource counters from the same offered-rate configuration, and, if no such point appears, report the measured point with maximum $o_i$ as a lower bound, breaking ties by lower $L_i$.

To evaluate the impact of system size on our key metrics, we varied $n$ from 10 to 100, distributing nodes evenly across 20 machines, with a maximum of 5 nodes per machine.
Thanks to the network's high bandwidth and low latency, this setup provides reliable results regarding message-passing behavior, with minimal and comparable CPU contention at 100 processes on 20 machines across all schemes.

To evaluate the impact of the batch size $B$ on our key metrics, we varied $B$ from 1 to 50.
For schemes that share multiple secrets by design (\ie \bingo, \hbacss, \light, and \goss), we used the described batching or packing mechanisms.
For \avss, we use a vectorized batching interface so that $B$ independent secrets are shared with $B$ independent sharing polynomials, while their private messages, acknowledgments, and transcript entries are grouped into one multi-secret execution.
This is the natural batched use of the transcript-based protocol, but it does not provide the same cryptographic amortization as packed or batch-native constructions.
Our \hbacss implementation follows the hbACSS0-style recovery path, and our \light implementation follows the beacon-based protocol's construction.
Because \haven and \honey lack native batching support, we evaluate them only with batch size 1.
Their per-secret overhead scales linearly with batch size, rendering large-batch experiments redundant and uninformative.
In our experiments, a secret is a 32-byte value, but larger secrets can be handled by batching multiple secrets together.

\bingo natively shares exactly $t+1$ secrets per batch for $n = 3t + 1$ parties.
Because its batch size is intrinsically tied to the system size, \bingo cannot be directly compared to other protocols in the previous scenarios.
We therefore conduct a separate set of experiments in which exactly $t+1$ secrets are shared at the same system size, enabling a fair comparison with the other batched protocols.
These experiments highlight the efficiency gains of \bingo's secret packing and provide a reference point for native batching in \light, \hbacss, and \goss, with \avss included through its vectorized multi-polynomial batching interface.
We evaluate this multi-secret stress test under both honest-dealer and faulty-dealer stress scenarios.

\section{Evaluations and results}
\label{sec:eval}

This section answers the five evaluation questions from Section~\ref{sec:method} by relating measured costs to the design dimensions of Section~\ref{sec:schemes}.
We restate them here to make the structure of the evaluation explicit.
\begin{enumerate}[left=0pt, itemsep=0pt,topsep=0pt,parsep=1pt]
	\item How large is the gap between optimistic and faulty-dealer stress executions?
	\item Which communication models dominate latency and throughput as the number of parties grows?
	\item Which cryptographic verification mechanisms become bottlenecks in practice?
	\item When does batching or packing amortize fixed costs, and when does it only increase latency?
	\item Which fault-handling strategies best preserve progress under faulty-dealer stress?
\end{enumerate}
For $B=1$, latency, bandwidth, message count, and CPU-time are per secret.
For batched experiments, the plotted bandwidth, message count, and CPU-time are per completed batch; the amortized per-secret cost is obtained by dividing by the number of secrets output by that batch.
For batched, packed, and vectorized multi-secret experiments, these metrics are per completed multi-secret sharing.
The output-rate plots in the peak-throughput experiments are normalized in secrets per second.

\subsection{RQ1: Gap between optimistic and faulty-dealer
	executions}

\begin{figure*}[!t]
	\centering
	\begin{subfigure}{\afourthofaline}
		\includegraphics[width=\linewidth]{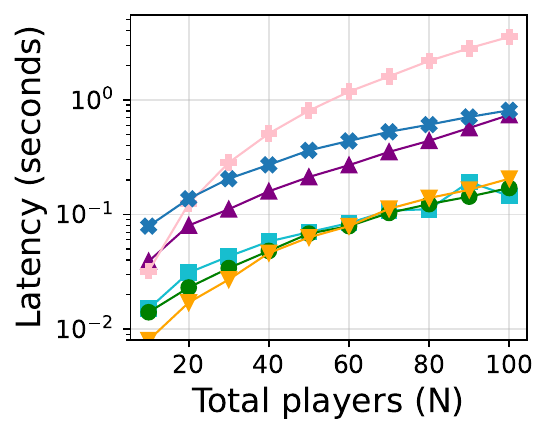}
		\caption{Latency}
	\end{subfigure}
	\begin{subfigure}{\afourthofaline}
		\includegraphics[width=\linewidth]{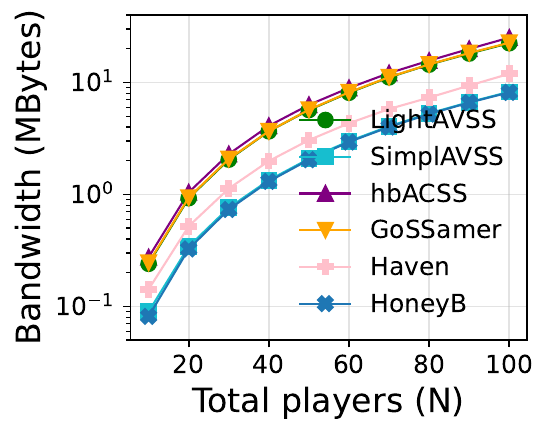}
		\caption{Bandwidth}
	\end{subfigure}
	\begin{subfigure}{\afourthofaline}
		\includegraphics[width=\linewidth]{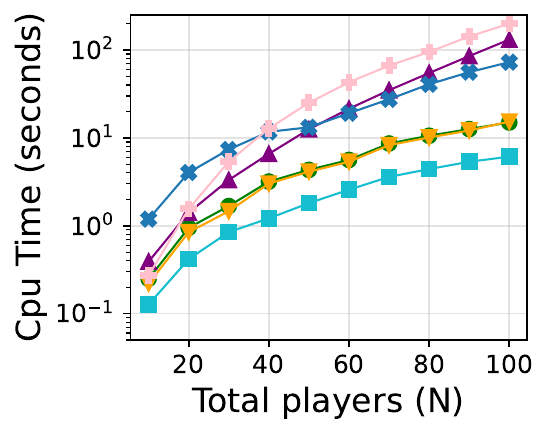}
		\caption{CPU-time}
	\end{subfigure}
	\begin{subfigure}{\afourthofaline}
		\includegraphics[width=\linewidth]{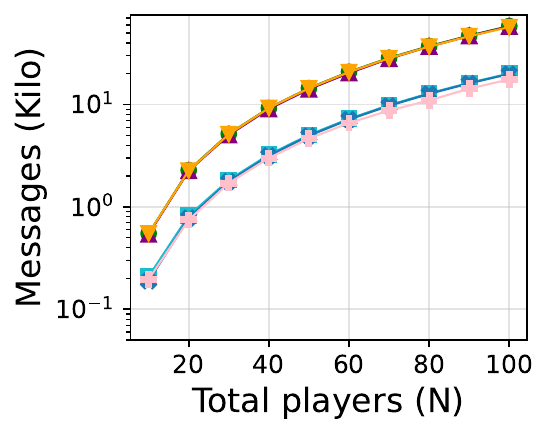}
		\caption{Messages exchanged}
	\end{subfigure}
	\caption{Scaling with $n$ for single-secret sharing under an honest dealer.}
	\label{fig:scale-single-honest}
\end{figure*}

\begin{figure*}[!t]
	\centering
	\begin{subfigure}{\afourthofaline}
		\includegraphics[width=\linewidth]{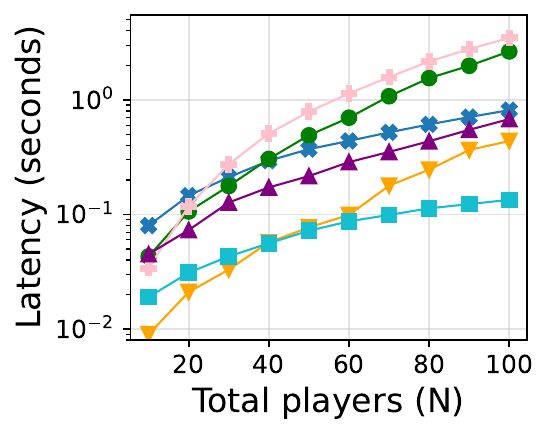}
		\caption{Latency}
	\end{subfigure}
	\begin{subfigure}{\afourthofaline}
		\includegraphics[width=\linewidth]{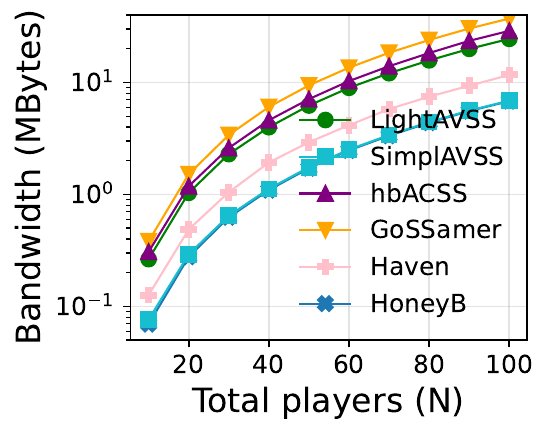}
		\caption{Bandwidth}
	\end{subfigure}
	\begin{subfigure}{\afourthofaline}
		\includegraphics[width=\linewidth]{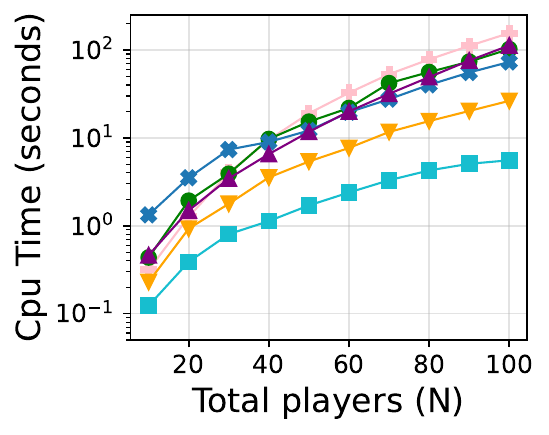}
		\caption{CPU-time}
	\end{subfigure}
	\begin{subfigure}{\afourthofaline}
		\includegraphics[width=\linewidth]{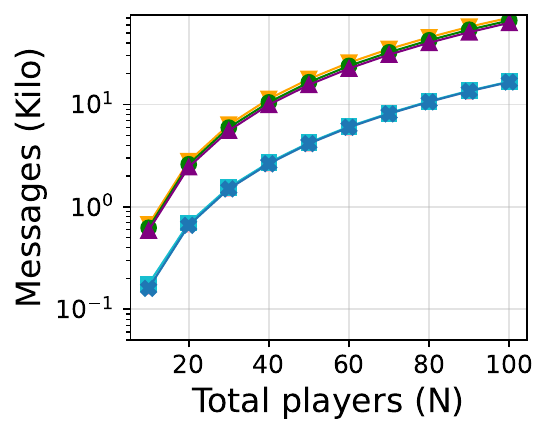}
		\caption{Messages exchanged}
	\end{subfigure}
	\caption{Scaling with $n$ for single-secret sharing under faulty-dealer stress.}
	\label{fig:scale-single-faulty}
\end{figure*}

Figures~\ref{fig:scale-single-honest} and~\ref{fig:scale-single-faulty} answer RQ1 for single-secret sharing by comparing the same protocols under an honest dealer and under faulty-dealer stress.

\begin{figure*}[!t]
	\centering
	\begin{subfigure}{\afourthofaline}
		\includegraphics[width=\linewidth]{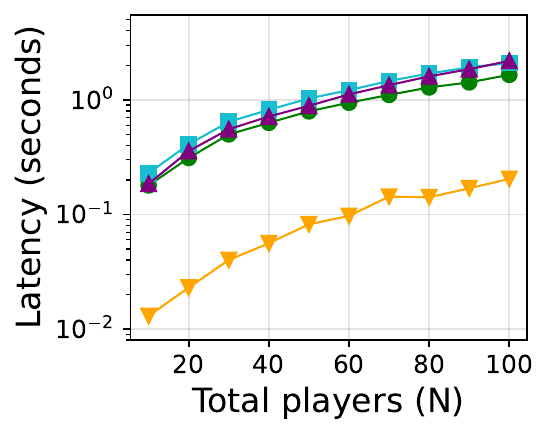}
		\caption{Latency}
	\end{subfigure}
	\begin{subfigure}{\afourthofaline}
		\includegraphics[width=\linewidth]{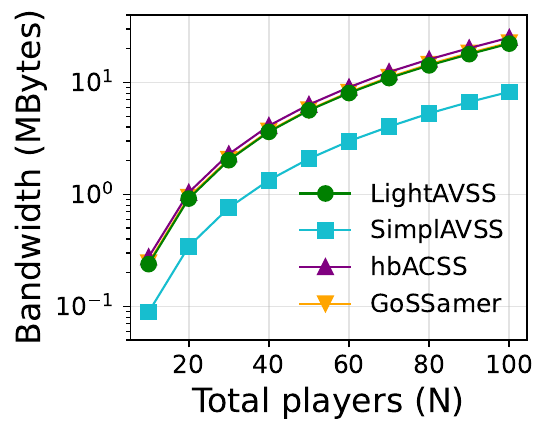}
		\caption{Bandwidth}
	\end{subfigure}
	\begin{subfigure}{\afourthofaline}
		\includegraphics[width=\linewidth]{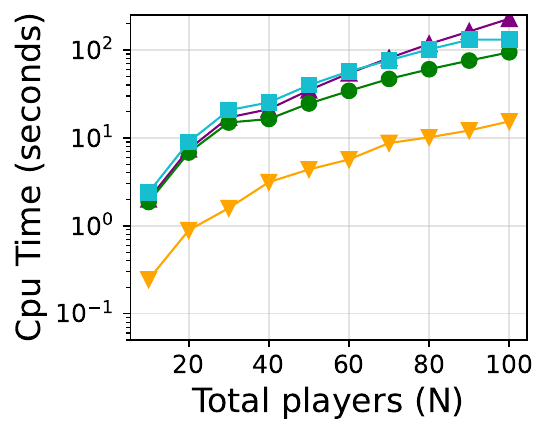}
		\caption{CPU-time}
	\end{subfigure}
	\begin{subfigure}{\afourthofaline}
		\includegraphics[width=\linewidth]{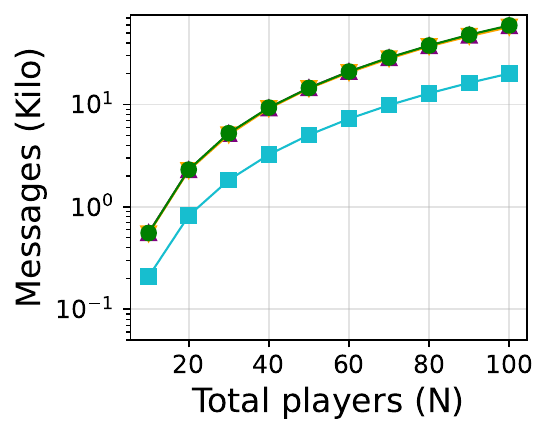}
		\caption{Messages exchanged}
	\end{subfigure}
	\caption{Scaling with $n$ for 50-secret batches under an honest dealer.}
	\label{fig:scale-batch50-honest}
\end{figure*}

\begin{figure*}[!t]
	\centering
	\begin{subfigure}{\afourthofaline}
		\includegraphics[width=\linewidth]{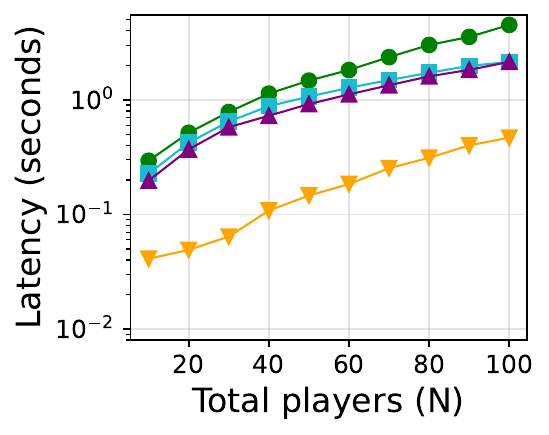}
		\caption{Latency}
	\end{subfigure}
	\begin{subfigure}{\afourthofaline}
		\includegraphics[width=\linewidth]{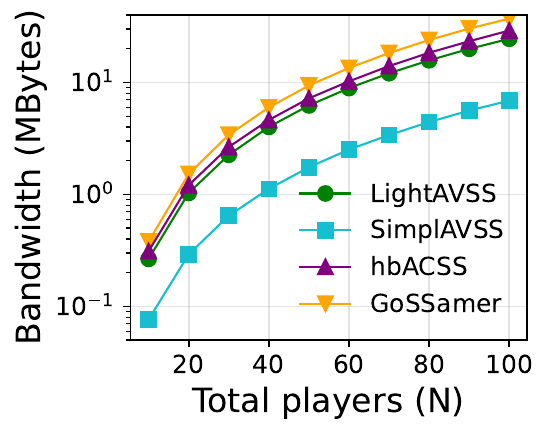}
		\caption{Bandwidth}
	\end{subfigure}
	\begin{subfigure}{\afourthofaline}
		\includegraphics[width=\linewidth]{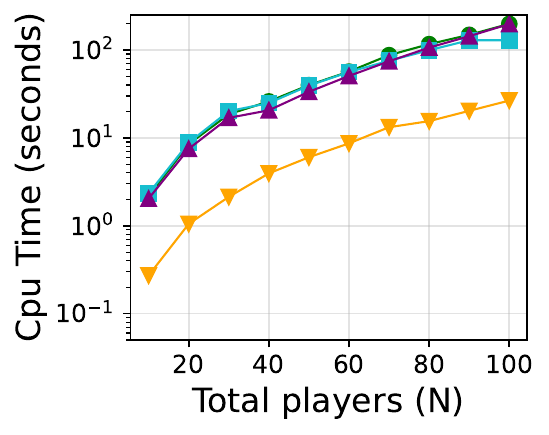}
		\caption{CPU-time}
	\end{subfigure}
	\begin{subfigure}{\afourthofaline}
		\includegraphics[width=\linewidth]{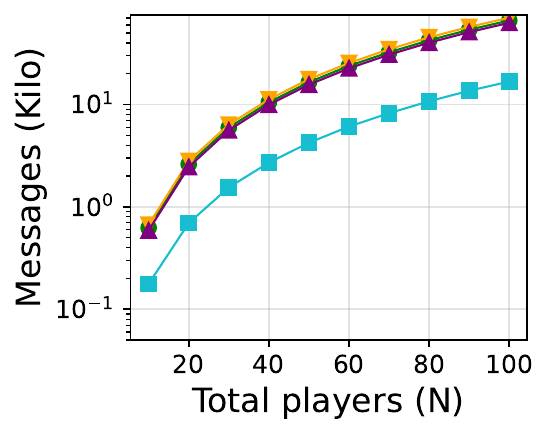}
		\caption{Messages exchanged}
	\end{subfigure}
	\caption{Scaling with $n$ for 50-secret batches under faulty-dealer stress.}
	\label{fig:scale-batch50-faulty}
\end{figure*}

\textit{Answer.}
The gap is small when the faulty path completes with cheap evidence or transcript completion, and large when an optimistic design moves complaint or recovery work onto the critical path.
Figures~\ref{fig:scale-single-honest}--\ref{fig:scale-batch50-faulty} show this separation directly.
For single-secret sharing at $n=100$, \light latency increases from 170\,ms under an honest dealer to 2.64\,s under faulty-dealer stress, while aggregate CPU-time rises from 15\,s to 102\,s.
\avss remains around 145\,ms and 134\,ms in the same two scenarios, with the lowest message and bandwidth costs among the implemented single-secret protocols.
\hbacss also keeps single-secret latency stable, but with substantially higher CPU-time than \avss because its batch-oriented verification machinery remains active.
For $B=50$ at $n=100$, this contrast persists: \light increases from 1.65\,s to 4.51\,s, whereas \goss increases from 204\,ms to 468\,ms.

Figures~\ref{fig:scale-batch50-honest} and~\ref{fig:scale-batch50-faulty} answer the same RQ for 50-secret multi-secret executions.

\subsection{RQ2: Dominating communications models for scalability}

Figures~\ref{fig:scale-single-honest}--\ref{fig:scale-batch50-faulty} answer RQ2 through their latency, bandwidth, and message-count panels as $n$ increases from 10 to 100.

\textit{Answer.}
No communication model dominates by message count alone; latency and throughput are dominated by where communication appears in the asynchronous critical path and by the verification it triggers.
In the honest single-secret experiment at $n=100$, \light, \hbacss, and \goss send about 57\,k--58\,k messages, whereas \avss, \honey, and \haven send about 18\,k--20\,k messages.
Despite this, \light and \goss remain latency-competitive, while \haven is the slowest protocol.
\haven spends about 199\,s of aggregate CPU-time at $n=100$, far above the lightweight hash- and symmetric-key paths.
\goss has the highest single-secret communication in the faulty-dealer stress scenario at $n=100$, with about 37\,MB and 70.5\,k messages, but remains below 0.5\,s latency.
\avss sends fewer messages and bytes, and is faster for single-secret sharing, but this advantage does not fully generalize to multi-secret workloads because vectorization groups transport messages while transcript evidence and commitment checks remain per secret.
For $B=50$, the vectorized \avss execution still sends fewer messages and bytes than the all-to-all batched designs, but its latency and CPU-time become comparable to \hbacss and worse than \goss.

\subsection{RQ3: Dominating cryptographic verification bottlenecks}

The CPU-time panels in Figures~\ref{fig:scale-single-honest}--\ref{fig:scale-batch50-faulty}, together with Figures~\ref{fig:bingo-style-honest} and~\ref{fig:bingo-style-faulty} for the packed setting, answer RQ3.

\textit{Answer.}
The main bottlenecks are public-key-heavy verification, polynomial-commitment checks, and adaptive packed verification.
Hash-based verification and original-evaluation propagation are substantially cheaper in the evaluated workloads.
\haven and \honey are dominated by heavier verification and public-key operations, which explains why their latency remains high even when their message counts are not the largest.
\hbacss incurs substantial CPU-time because batch polynomial-commitment verification and recovery checks are present even when the batch is small.
\bingo shows the same effect in the packed setting.
Its communication is lower than \light, \hbacss, and \goss at $n=100$, but its CPU-time is much higher than \goss and comparable to the most expensive batched designs.
\goss keeps CPU-time low across honest-dealer and faulty-dealer stress batched executions because its core path uses hash-based verification and original-evaluation propagation rather than homomorphic commitments.
\light also benefits from lightweight verification in the honest path, but the advantage largely disappears under faulty-dealer stress.
\avss is efficient for a single secret, yet its CPU-time grows in vectorized batches because public transcript processing scales with the number of independently shared secrets.

\subsection{RQ4: Batching and packing}

\begin{figure}[!t]
	\centering
	\begin{subfigure}{\halfaline}
		\includegraphics[width=\linewidth]{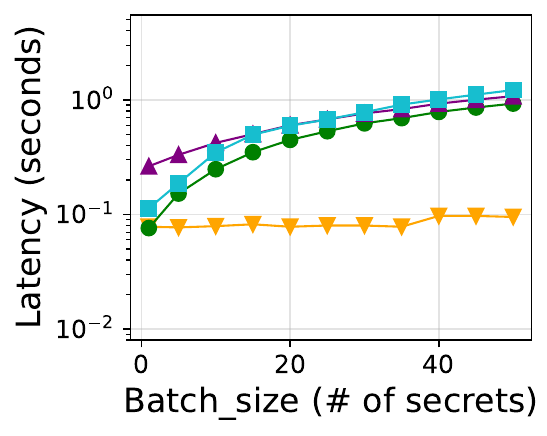}
		\caption{Latency}
	\end{subfigure}
	\begin{subfigure}{\halfaline}
		\includegraphics[width=\linewidth]{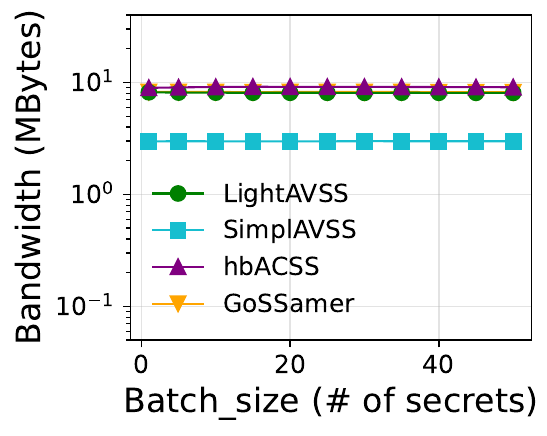}
		\caption{Bandwidth}
	\end{subfigure}
	\begin{subfigure}{\halfaline}
		\includegraphics[width=\linewidth]{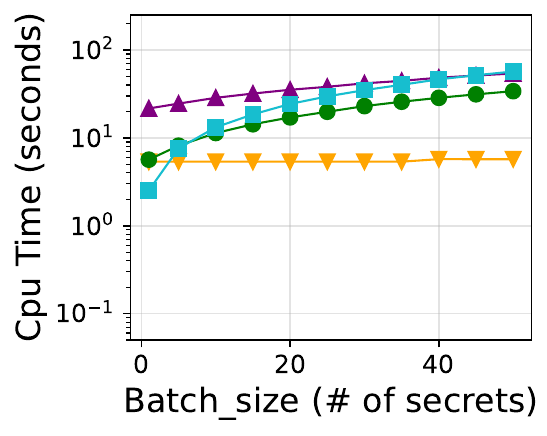}
		\caption{CPU-time}
	\end{subfigure}
	\begin{subfigure}{\halfaline}
		\includegraphics[width=\linewidth]{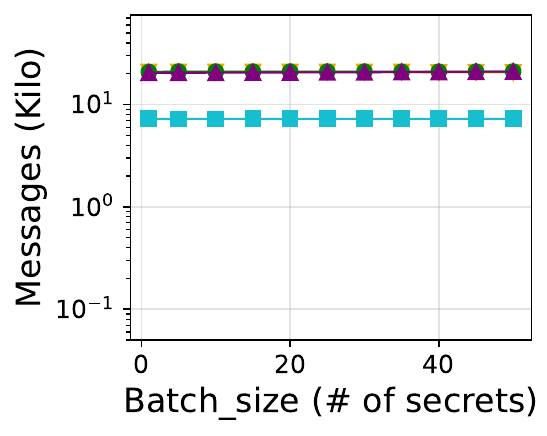}
		\caption{Messages exchanged}
	\end{subfigure}
	\caption{Impact of batch size at $n=60$ under an honest dealer.}
	\label{fig:batch-size-honest}
\end{figure}

\begin{figure}[!t]
	\centering
	\begin{subfigure}{\halfaline}
		\includegraphics[width=\linewidth]{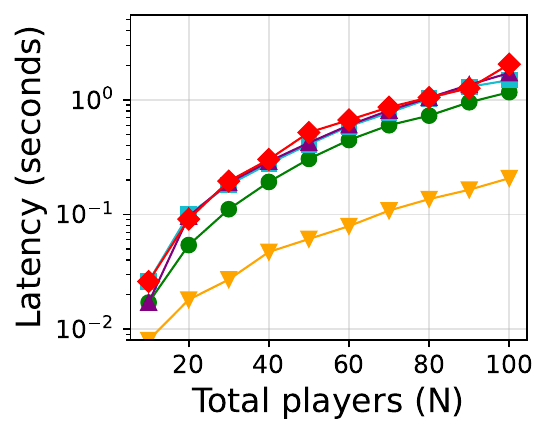}
		\caption{Latency}
	\end{subfigure}
	\begin{subfigure}{\halfaline}
		\includegraphics[width=\linewidth]{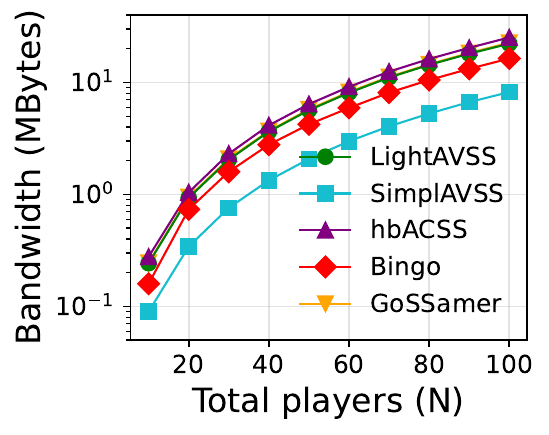}
		\caption{Bandwidth}
	\end{subfigure}
	\begin{subfigure}{\halfaline}
		\includegraphics[width=\linewidth]{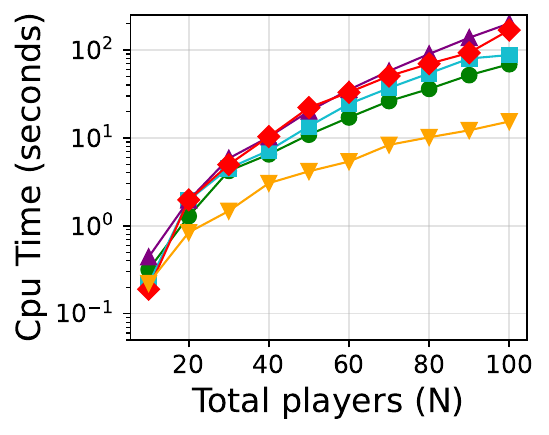}
		\caption{CPU-time}
	\end{subfigure}
	\begin{subfigure}{\halfaline}
		\includegraphics[width=\linewidth]{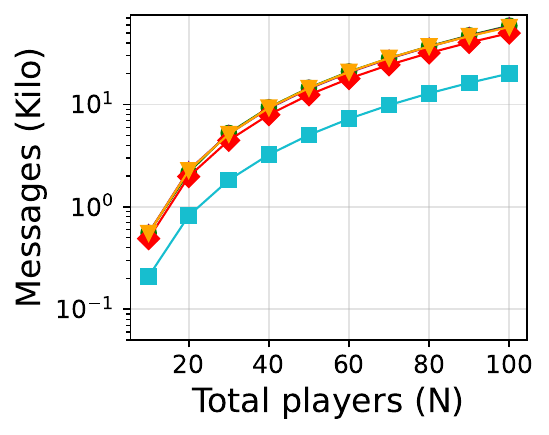}
		\caption{Messages exchanged}
	\end{subfigure}
	\caption{Bingo-style multi-secret sharing under an honest dealer, with $B=t+1$ and $n=3t+1$.}
	\label{fig:bingo-style-honest}
\end{figure}

\begin{figure}[!t]
	\centering
	\begin{subfigure}{\halfaline}
		\includegraphics[width=\linewidth]{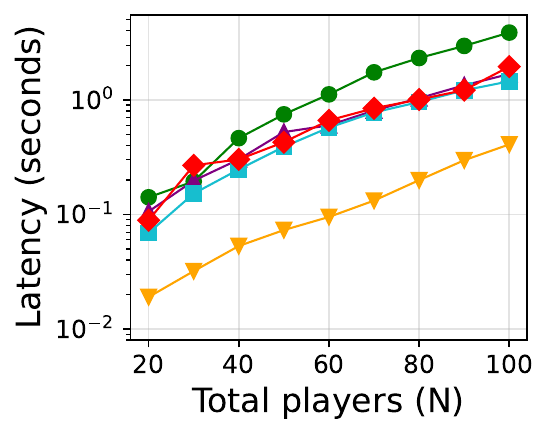}
		\caption{Latency}
	\end{subfigure}
	\begin{subfigure}{\halfaline}
		\includegraphics[width=\linewidth]{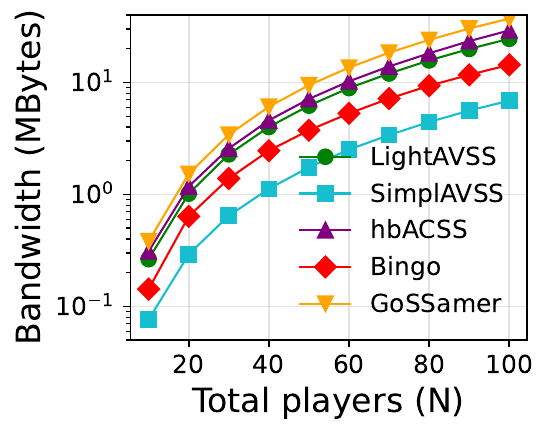}
		\caption{Bandwidth}
	\end{subfigure}
	\begin{subfigure}{\halfaline}
		\includegraphics[width=\linewidth]{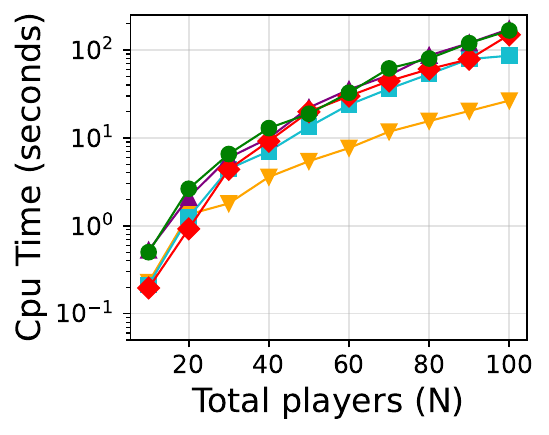}
		\caption{CPU-time}
	\end{subfigure}
	\begin{subfigure}{\halfaline}
		\includegraphics[width=\linewidth]{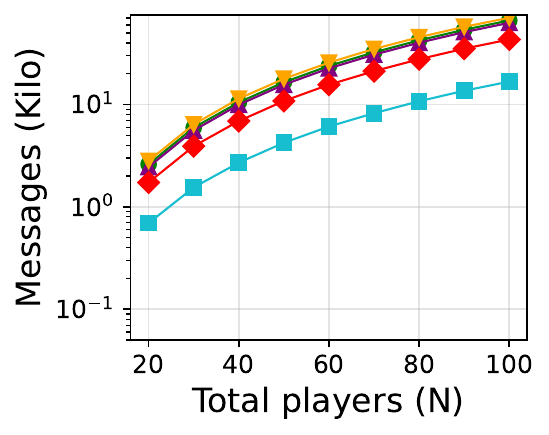}
		\caption{Messages exchanged}
	\end{subfigure}
	\caption{Bingo-style multi-secret sharing under faulty-dealer stress, with $B=t+1$ and $n=3t+1$.}
	\label{fig:bingo-style-faulty}
\end{figure}

Figure~\ref{fig:batch-size-honest} answers RQ4 for fixed $n=60$ by varying the number of secrets per execution.
Figures~\ref{fig:bingo-style-honest} and~\ref{fig:bingo-style-faulty} answer the same RQ for the packed $B=t+1$ setting under honest-dealer and faulty-dealer stress executions.

\textit{Answer.}
Batching amortizes fixed costs only when communication and verification are shared across secrets; otherwise it mostly groups executions and increases batch latency or CPU-time.
Figure~\ref{fig:batch-size-honest} shows this distinction.
At $n=60$, the message and byte counts of \light, \hbacss, and \goss are nearly flat as $B$ grows from 1 to 50, while the number of secrets output by each execution increases.
Their per-secret communication therefore decreases with $B$, even when the plotted per-batch cost remains constant.
\avss follows the same pattern: grouping messages keeps byte and message counts low, but latency and CPU-time still grow with $B$ because transcript evidence remains per secret.
At $n=60$ and $B=50$, \goss completes in about 95\,ms with about 5.7\,s of aggregate CPU-time, while \light, \hbacss, and vectorized \avss range from about 0.93\,s to 1.22\,s and from about 34\,s to 57\,s of aggregate CPU-time.
The trade-off is that \goss does not minimize bytes or messages; it minimizes critical-path computation and latency for a batch when sufficient bandwidth is available.

Figures~\ref{fig:bingo-style-honest} and~\ref{fig:bingo-style-faulty} isolate the packed setting by running native batched or packed schemes with $B=t+1$, together with vectorized \avss.
\bingo reduces communication compared with \light, \hbacss, and \goss, especially under faulty-dealer stress, but this saving comes with high CPU-time from adaptive-security and packed-verification machinery.
At $n=100$ under faulty-dealer stress, \bingo uses about 14.4\,MB and 43\,k messages, while \goss uses about 37.2\,MB and 70.5\,k messages.
However, \goss remains below 0.5\,s latency and about 26.5\,s CPU-time, whereas \bingo is near 1.95\,s latency and 149\,s CPU-time.

\subsection{RQ5: Which fault-handling strategies best preserve progress with a faulty dealer?}

Figures~\ref{fig:scale-single-faulty},~\ref{fig:scale-batch50-faulty}, and~\ref{fig:bingo-style-faulty} answer RQ5 at low offered load, while Figures~\ref{fig:peak-faulty} and~\ref{fig:peak-bingo-style} show how the same fault-handling paths behave near saturation.

\textit{Answer.}
Fault-handling strategies best preserve progress when they complete with cheap, locally checkable evidence and avoid commitment-heavy repair on the critical path.
\avss uses transcript-based completion, so injected dealer faults do not induce a large communication or CPU penalty in the single-secret case.
\hbacss uses explicit complaint and interpolation machinery; it preserves completion and keeps latency predictable, but its CPU-time remains high in both honest-dealer and faulty-dealer stress executions.
\light uses an optimistic path and therefore suffers the largest recovery penalty, especially for batched executions.
\goss preserves recovery progress by propagating original authenticated evaluations instead of invoking commitment-heavy repair.
\bingo avoids the collapse seen in \light, but its packed and adaptively secure verification path makes it CPU-bound relative to \goss and \avss.
\haven and \honey are stable in the sense that injected dealer faults do not radically change their ranking, but their baseline cost is already dominated by high-threshold and public-verification machinery.

\subsection{Peak-throughput saturation}

\begin{figure}[!t]
	\centering
	\begin{subfigure}{\halfaline}
		\includegraphics[width=\linewidth]{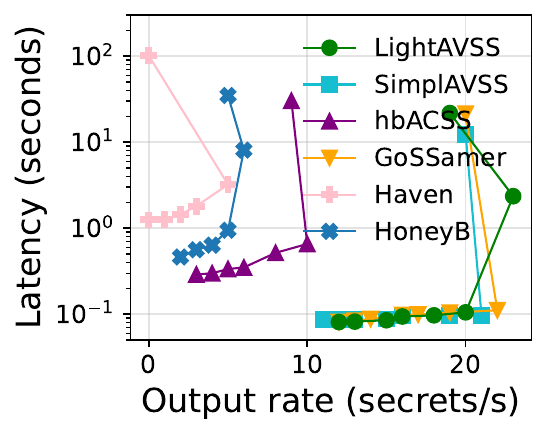}
		\caption{Single-secret sharing}
	\end{subfigure}
	\begin{subfigure}{\halfaline}
		\includegraphics[width=\linewidth]{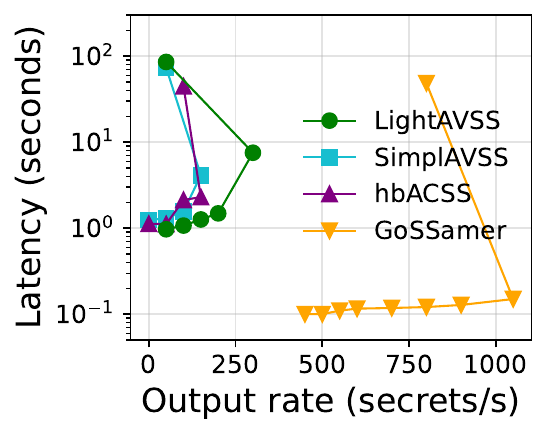}
		\caption{Batch size $B=50$}
	\end{subfigure}
	\caption{Peak performance under an honest dealer, $n=60$.}
	\label{fig:peak-honest}
\end{figure}

\begin{figure}[!t]
	\centering
	\begin{subfigure}{\halfaline}
		\includegraphics[width=\linewidth]{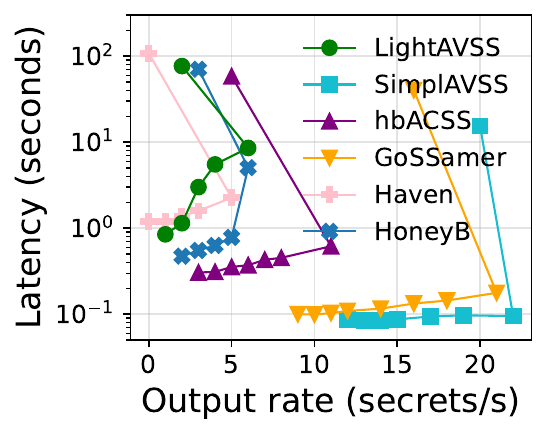}
		\caption{Single-secret sharing}
	\end{subfigure}
	\begin{subfigure}{\halfaline}
		\includegraphics[width=\linewidth]{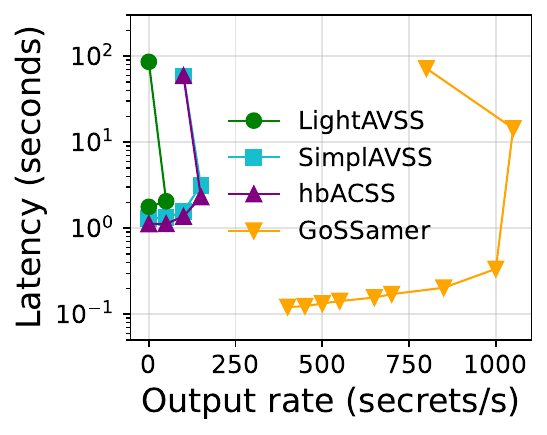}
		\caption{Batch size $B=50$}
	\end{subfigure}
	\caption{Peak performance under faulty-dealer stress, $n=60$.}
	\label{fig:peak-faulty}
\end{figure}

\begin{figure}[!t]
	\centering
	\begin{subfigure}{\halfaline}
		\includegraphics[width=\linewidth]{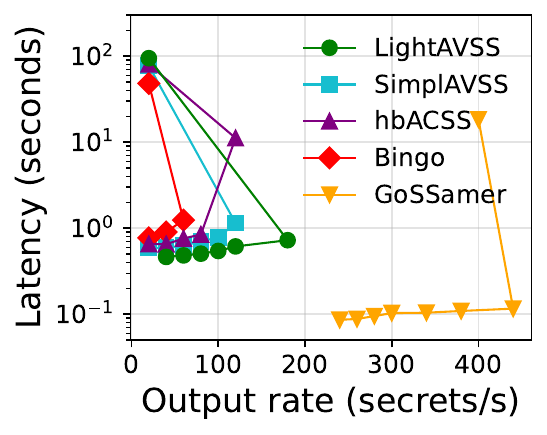}
		\caption{Honest dealer}
	\end{subfigure}
	\begin{subfigure}{\halfaline}
		\centering
		\includegraphics[width=\linewidth]{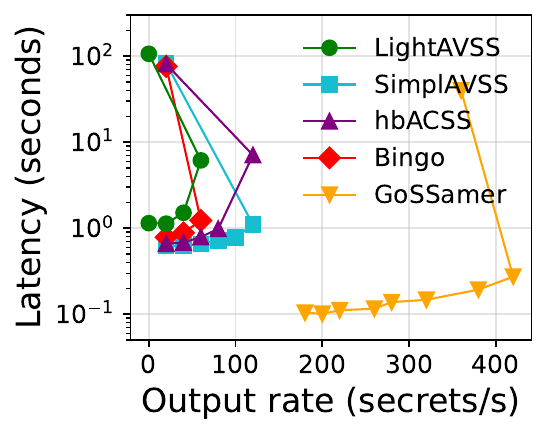}
		\caption{Faulty-dealer stress}
	\end{subfigure}
	\caption{Peak performance for the packed setting, $n=61$, $B=t+1=21$.}
	\label{fig:peak-bingo-style}
\end{figure}

Figures~\ref{fig:peak-honest}--\ref{fig:peak-bingo-style} complement the previous per-execution measurements by reporting output rate and 99th-percentile latency as the offered load approaches saturation.
They answer the throughput component of RQ2 and give a load-stress view of the batching and fault-handling mechanisms studied in RQ4 and RQ5.
Table~\ref{tab:peak-knee-costs} reports the achieved output rate, average completion latency, and resource counters measured at the selected saturation knee.
In the table, throughput is reported in secrets per second, latency in seconds, messages in thousands, bandwidth in MB, and CPU-time in aggregate CPU-seconds.

\begin{table}[!t]
     \centering
     \caption{Performance and costs at the peak-throughput knee.}
     \label{tab:peak-knee-costs}
     \setlength{\tabcolsep}{2pt}
     \resizebox{\columnwidth}{!}{%
          \begin{tabular}{@{}lllrrrrr@{}}
               \toprule
               Case & Params                              & Prot.   & Thr.  & Lat.  & Msg.  & BW    & CPU    \\
               \midrule
               \multirow{10}{*}{Honest}
                    & \multirow{6}{*}{\makecell[l]{$B=1$                                                     \\$n=60$}}
                    & \haven                              & 5       & 2.581 & 6.1   & 4.12  & 702.9          \\
                    &                                     & \hbacss & 10    & 0.469 & 20.5  & 9.07  & 116.4  \\
                    &                                     & \honey  & 6     & 3.518 & 7.1   & 2.93  & 819.6  \\
                    &                                     & \light  & 23    & 0.151 & 20.8  & 8.10  & 15.9   \\
                    &                                     & \avss   & 21    & 0.090 & 7.2   & 2.98  & 4.9    \\
                    &                                     & \goss   & 22    & 0.100 & 20.7  & 8.19  & 12.2   \\
               \cmidrule{2-8}
                    & \multirow{4}{*}{\makecell[l]{$B=50$                                                    \\$n=60$}}
                    & \hbacss                             & 150     & 1.825 & 21.0  & 9.03  & 418.9          \\
                    &                                     & \light  & 300   & 4.789 & 21.3  & 8.01  & 1523.8 \\
                    &                                     & \avss   & 150   & 3.275 & 7.3   & 2.99  & 980.3  \\
                    &                                     & \goss   & 1050  & 0.131 & 20.6  & 8.19  & 16.7   \\
               \midrule
               \midrule
               \multirow{10}{*}{Faulty}
                    & \multirow{6}{*}{\makecell[l]{$B=1$                                                     \\$n=60$}}
                    & \haven                              & 5       & 1.787 & 6.0   & 4.09  & 369.9          \\
                    &                                     & \hbacss & 11    & 0.517 & 22.6  & 10.32 & 143.6  \\
                    &                                     & \honey  & 6     & 2.639 & 6.0   & 2.47  & 806.5  \\
                    &                                     & \light  & 6     & 7.331 & 23.8  & 8.80  & 561.3  \\
                    &                                     & \avss   & 22    & 0.089 & 6.1   & 2.51  & 4.7    \\
                    &                                     & \goss   & 21    & 0.151 & 25.6  & 13.52 & 24.1   \\
               \cmidrule{2-8}
                    & \multirow{4}{*}{\makecell[l]{$B=50$                                                    \\$n=60$}}
                    & \hbacss                             & 150     & 1.944 & 22.6  & 10.43 & 486.5          \\
                    &                                     & \light  & 50    & 1.850 & 23.9  & 8.86  & 104.2  \\
                    &                                     & \avss   & 150   & 2.649 & 6.1   & 2.51  & 723.8  \\
                    &                                     & \goss   & 1000  & 0.165 & 25.5  & 13.50 & 148.6  \\
               \midrule
               \midrule
               \multirow{5}{*}{Bingo honest}
                    & \multirow{5}{*}{\makecell[l]{$B=21$                                                    \\$n=61$}}
                    & \bingo                              & 60      & 1.139 & 18.5  & 6.19  & 149.0          \\
                    &                                     & \hbacss & 120   & 6.251 & 21.5  & 9.04  & 590.6  \\
                    &                                     & \light  & 180   & 0.635 & 21.6  & 8.39  & 108.1  \\
                    &                                     & \avss   & 120   & 0.907 & 7.5   & 3.09  & 195.7  \\
                    &                                     & \goss   & 440   & 0.104 & 21.4  & 8.49  & 13.0   \\
               \cmidrule{1-8}
               \multirow{5}{*}{Bingo faulty}
                    & \multirow{5}{*}{\makecell[l]{$B=21$                                                    \\$n=61$}}
                    & \bingo                              & 60      & 1.108 & 16.1  & 5.45  & 136.6          \\
                    &                                     & \hbacss & 120   & 3.337 & 23.3  & 10.72 & 894.5  \\
                    &                                     & \light  & 60    & 4.400 & 24.6  & 9.21  & 422.4  \\
                    &                                     & \avss   & 120   & 0.900 & 6.3   & 2.58  & 192.2  \\
                    &                                     & \goss   & 420   & 0.198 & 26.4  & 13.96 & 35.7   \\
               \bottomrule
          \end{tabular}
     }
\end{table}

For $B=1$, Figures~\ref{fig:peak-honest} and~\ref{fig:peak-faulty} show that the saturation ranking is governed more by critical-path work than by raw message count.
\avss, \light, and \goss sustain similar output rates under an honest dealer, while \haven and \honey reach lower rates despite moderate communication because their verification paths are more expensive.
Under faulty-dealer stress, \avss remains stable and \goss keeps comparable throughput while spending more messages and bytes, whereas \light loses most of its single-secret throughput because complaint and recovery work move onto the critical path.

For $B=50$, the same figures show whether batching actually amortizes the limiting resource.
\goss sustains around one thousand secrets per second with sub-200\,ms average latency at the knee in both honest-dealer and faulty-dealer peak experiments because its batched path keeps CPU-time low and avoids commitment-heavy recovery.
The cost is visible in communication, especially under faulty-dealer stress.
By contrast, \hbacss and vectorized \avss reach lower batch throughput because polynomial-commitment checks or transcript processing remain expensive, while \light is highly sensitive to whether the recovery path is activated.
Figure~\ref{fig:peak-bingo-style} shows the same design tension in the $B=t+1$ packed setting.
\bingo reduces network cost relative to the all-to-all batched designs, but its packed and adaptive-verification machinery limits peak throughput.
\goss sustains a higher knee by keeping CPU-time low while paying more messages and bytes.
The peak results therefore refine the low-load measurements: a protocol that is efficient for isolated sharing executions does not necessarily sustain the highest load, and a protocol that sends more data can still reach a higher knee if it avoids CPU-heavy verification on the asynchronous critical path.

\subsection{Synthesis}

The evaluation supports the design-space sieve.
Optimistic execution explains \light's honest-dealer advantage and its large faulty-dealer penalty; transcript-based completion explains \avss's single-secret efficiency and batching limits; batch-oriented recovery explains \hbacss's predictable but CPU-heavy behavior; original-evaluation propagation explains \goss's batched recovery behavior; and packed bivariate sharing explains \bingo's communication savings and verification cost.
Protocol choice is therefore multi-objective, depending on whether the deployment bottleneck is latency, CPU-time, bandwidth, fault recovery, or adaptive-security requirements.

\section{Discussion}
\label{sec:discussion}

This section discusses the scope and implications of our findings.
We first identify threats to the validity and generalizability of the evaluation, then translate the observed design-to-cost trade-offs into guidance for practitioners, and finally outline research directions suggested by the results.
\begin{table*}[!ht]
    \caption{Guidelines for recommending AVSS and ACSS protocols based on one-shot sharing priorities.}
    \label{tab:guidelines}
    \centering
    \small{
        \begin{tabular}{
                @{}
                P{0.2\textwidth}
                P{0.1\textwidth}
                P{0.68\textwidth}
                @{}
            }
            \toprule
            \textbf{Priority}
             & \textbf{Rec. scheme}
             & \textbf{Empirical Justification}                                                                                               \\
            \midrule
            Low latency (single secret)
             & \avss
             & Lowest latency and communication in the single-secret experiments under both honest-dealer and faulty-dealer stress scenarios                                               \\
            \midrule
            Low latency and CPU for batches
             & \goss
             & Best batched latency and CPU-time in the current experiments; recovery mainly increases bytes and messages                                    \\
            \midrule
            Predictable batch recovery
             & \hbacss
             & Stable latency under faulty-dealer stress, but with higher verification cost than lightweight hash-based designs                                    \\
            \midrule
            Linear batched ACSS
             & \goss
             & Amortizes communication over large batches using lightweight verification and preserves the same asymptotic bound in recovery \\
            \midrule
            Low communication volume
             & \avss \newline \bingo
             & \avss minimizes communication for single-secret and vectorized multi-secret sharing, while \bingo reduces communication in the packed $t+1$ setting
            \\
            \midrule
            Packed $t+1$ sharing
             & \bingo
             & Shares $t+1$ secrets per instance and supports adaptive security, with lower communication but higher verification cost                           \\
            \midrule
            Low cryptographic cost
             & \light \newline \goss
             & Rely on lightweight primitives such as hashes, Merkle trees, and hash-based degree checks; no public-key crypto or zero-knowledge proofs in the core verification path \\
            \bottomrule
        \end{tabular}
    }
\end{table*}

\subsection{Threats to Validity}
\label{sec:threats}

Our common framework controls the network stack, serialization, process placement, logging, and shared subprotocol implementations, but it cannot eliminate all implementation bias.
The protocols differ in maturity, cryptographic assumptions, security guarantees, and opportunities for optimization, and our implementations may not capture every optimization available to protocol-specific codebases.
We mitigate this threat by following the original specifications, using common cryptographic libraries and infrastructure where possible, and interpreting the results only after filtering protocols by their required security envelope.
The selected protocols are representative design points rather than an exhaustive implementation of the corpus, so the conclusions apply primarily to the mechanisms and one-shot sharing interfaces evaluated here.

Our faulty-dealer experiment is a controlled stress scenario that activates detection, complaint, or recovery paths; it is not an exhaustive Byzantine worst case and does not cover all equivocation patterns, message schedules, or adaptive corruptions.
Similarly, Grid'5000 provides a homogeneous, low-latency environment that helps isolate intrinsic protocol costs but does not reproduce geo-distributed WAN conditions or heterogeneous deployments.
The reported averages and peak-throughput knees should therefore be read as comparative measurements under the stated workloads and placement policy, not as universal deployment predictions.

\subsection{Implications for Practitioners}
\label{sec:practitioners}

Practitioners should use the results as a deployment filter rather than as a global ranking.
First, eliminate protocols whose adversary model, privacy threshold, setup model, or trust assumptions do not match the application.
Then choose according to the deployment bottleneck.
If most executions are expected to remain fault-free, an optimistic common path can be worth the recovery penalty; if faulty-dealer completion is expected to be on the critical path, transcript completion or recovery-native designs are safer.
If the workload is dominated by isolated secrets, avoid paying batch-native proof or dispersal costs; if it consists of bursts of many secrets, prefer designs whose dissemination and verification amortize across the batch.
Vectorized transcript batching is useful for grouping transport work, but should not be treated as the same amortization as packing or native batching.
Finally, when high-threshold privacy, public verifiability, or adaptive security is mandatory, those guarantees should dominate the performance ranking.
These performance recommendations apply only after the deployment's required adversary model, privacy threshold, setup assumptions, and trust assumptions have been satisfied.
Table~\ref{tab:guidelines} summarizes these trade-offs and provides concrete guidance for selecting an AVSS or ACSS protocol based on deployment conditions, fault assumptions, and performance requirements.

\subsection{Future Research Directions}
\label{sec:future}

The evaluation suggests several research directions for AVSS and ACSS protocols that are not captured by optimizing a single asymptotic metric.
The central lesson is that completion, verification, batching, and recovery interact.
A protocol can send more data but complete faster if recovery avoids expensive work on the asynchronous critical path, while a protocol with low communication can still become CPU-bound under batching or adaptive security.

\textbf{Critical-path-aware recovery.}
Faulty-dealer stress executions show that the cost of recovery depends less on whether additional data is exchanged than on whether verification and interpolation are placed on the latency-critical path.
A useful direction is to design recovery mechanisms that preserve ACSS-style completion while exposing only cheap, locally checkable evidence to the critical path, deferring heavier repair or audit work whenever possible.
This raises a precise protocol question.
What is the minimum evidence that honest parties must exchange to guarantee completion, agreement on the shared value, and later reconstructability under asynchrony?
Answering it would help separate unavoidable recovery cost from artifacts of current complaint, transcript, or interpolation mechanisms.

\textbf{Adaptive batching and packing.}
Our results show that batching is beneficial only when communication and verification costs amortize cleanly, and that packing can reduce communication while increasing CPU-time.
Future protocols should therefore treat batch size as a systems parameter rather than a fixed construction parameter.
Open problems include adaptive batching policies that react to workload pressure and detected faults, hybrids that combine packed encodings with lightweight verification, and one-shot interfaces that support both latency-sensitive single secrets and throughput-oriented batches without switching protocols.
Such designs would be especially useful for MPC, DKG, and randomness-generation workloads, where the number of secrets available for batching is often bursty rather than fixed.

\textbf{Scalable dissemination under AVSS/ACSS semantics.}
The communication results suggest that simply reducing message count is insufficient; scalable dissemination must also preserve asynchronous completion, verifiability, and recovery semantics.
Clustered~\cite{yahyaoui2024tolerating, amir2008steward}, probabilistic~\cite{guerraoui2019scalable}, and optimistic~\cite{shrestha2025optimistic} communication techniques provide useful starting points, but applying them to AVSS and ACSS is not a mechanical substitution for reliable broadcast or secure message distribution.
The main challenge is to reduce all-to-all pressure without introducing hidden synchrony assumptions, weakening the corruption threshold, or making recovery depend on small committees that can be selectively targeted.
A rigorous direction is therefore to characterize which dissemination structures can safely carry shares, commitments, and recovery evidence in a fully asynchronous adversarial model.

\textbf{Verification primitives for many recipients.}
Several evaluated protocols are limited by verification cost rather than by raw bytes or messages.
Recent progress on one-to-many provers~\cite{zhang2022polynomial}, transparent polynomial commitments such as PolyFRI~\cite{zhang2024fast}, and high-throughput proof systems such as HyperPlonk~\cite{chen2023hyperplonk} suggests an opportunity to design commitments specifically for asynchronous multi-recipient sharing.
The relevant target is not only shorter proofs, but proofs that support many evaluations, batched recipient checks, recovery openings, and cheap aggregation across concurrent sharing instances.
An AVSS/ACSS-oriented commitment interface should make explicit which operations are on the dealer's path, on each recipient's verification path, and on the recovery path.

\textbf{Hardware-assisted sharing with explicit fallback.}
Trusted Execution Environments can simplify Byzantine protocols and reduce fault-tolerance costs in related settings~\cite{decouchant2022damysus, gupta2023dissecting}, and they may also reduce the need for some heavyweight cryptographic checks in AVSS and ACSS.
However, TEEs change the threat model rather than merely improving performance.
A scientifically sound direction is hybrid sharing in which hardware assistance accelerates the common path, but compromise, side channels, rollback, or attestation failure trigger a cryptographic recovery path with well-defined guarantees.
This would make the performance benefit of TEEs measurable without making the security argument depend entirely on trusted hardware.

\section{Conclusion}
\label{sec:conclusion}

This paper presents a design-space analysis and systems evaluation of modern one-shot AVSS and ACSS constructions.
Through a unified taxonomy and an apples-to-apples evaluation of seven representative designs, we show how performance and cost trade-offs are primarily driven by a small set of architectural choices, including communication model, cryptographic verification, recovery strategy, batching or packing, and optimistic versus robust-first execution.
Our results highlight that no single protocol dominates across workloads.
Designs optimized for simplicity or optimistic settings scale poorly under large batches or faults, while batch-oriented and packed protocols improve scalability only when verification and recovery work amortize cleanly.
These findings offer practical guidance for protocol selection and inform the design of future asynchronous verifiable and complete sharing protocols.
All code, experimental data, and corpus metadata are released as open artifacts to support reproducibility and further investigation.

\bibliographystyle{IEEEtran}
\bibliography{references.bib}

\end{document}